\documentclass[journal]{IEEEtran}
\ifCLASSINFOpdf
\else
\fi

\usepackage{graphicx}
\usepackage{amsfonts}
\usepackage{caption}
\usepackage[labelformat=simple]{subcaption}
\usepackage{float}
\usepackage{amssymb}
\usepackage{enumitem}
\usepackage{algorithm} 
\usepackage{algorithmic}
\usepackage[algo2e]{algorithm2e}
\usepackage{amsmath,amsthm,mathtools}
\usepackage{url}
\usepackage{stmaryrd}
\usepackage{cite}
\usepackage{multirow}
\usepackage[table,xcdraw]{xcolor}
\usepackage{url}
\usepackage{ragged2e}
\usepackage{epstopdf}

\newcommand{\etal}{\textit{et al}. }
\newcommand{\ie}{\textit{i}.\textit{e}., }

\setlist{nolistsep}

\DeclareMathOperator{\diag}{\operatorname{diag}}
\DeclareMathOperator{\vectorized}{\operatorname{vec}}

\DeclareMathOperator{\tr}{\operatorname{tr}}

\newtheorem{theorem}{Theorem}

\newtheorem{definition}{Definition}

\begin{document}
\bstctlcite{IEEEexample:BSTcontrol}
\title{Reconstruction of Time-varying Graph Signals via Sobolev Smoothness}
%Graph Signal Processing for the Estimation of New COVID-19 Cases}

\author{Jhony~H.~Giraldo,~Arif~Mahmood,~Belmar~Garcia-Garcia,~Dorina~Thanou, and~Thierry~Bouwmans% <-this % stops a space
%\author{Jhony~H.~Giraldo,~Arif~Mahmood,~Dorina~Thanou, and~Thierry~Bouwmans% <-this % stops a space
\thanks{Jhony H. Giraldo, Belmar Garcia-Garcia, and Thierry Bouwmans are with the laboratoire MIA, Mathématiques, Image et Applications, La Rochelle Université, 17000 La Rochelle, France e-mail: jgiral01@univ-lr.fr, belmar\_2g@hotmail.com, tbouwman@univ-lr.fr.}% <-this % stops a space
\thanks{Arif Mahmood is with the Department of Computer Science, Information Technology University, 48730 Lahore, Pakistan e-mail:arif.mahmood@itu.edu.pk.}% <-this % stops a space
\thanks{Dorina Thanou is with the Center for Intelligent Systems, EPFL, 1015 Lausanne, Switzerland e-mail: dorina.thanou@epfl.ch.}% <-this % stops a space
\thanks{The authors thank the support provided by CONACYT Mexico under the grant number 770750.}% <-this % stops a space
}


%\markboth{IEEE Transactions on Signal and Information Processing over Networks}%
\markboth{Signal Processing}%
{Giraldo \MakeLowercase{\textit{et al.}}: Reconstruction of Time-varying Graph Signals via Sobolev Smoothness}



\maketitle

\begin{abstract}
    Graph Signal Processing (GSP) is an emerging research field that  extends the concepts of digital signal processing to graphs.
    GSP has numerous applications in different areas such as sensor networks, machine learning, and image processing.
    The sampling and reconstruction of static graph signals have played a central role in GSP.
    However, many real-world graph signals are  inherently time-varying and 
    the smoothness of the temporal differences of such graph signals may be used as a prior assumption.
    In the current work, we assume that the temporal differences of graph signals are smooth, and we introduce a novel algorithm based on the extension of a Sobolev smoothness function for the reconstruction of time-varying graph signals from discrete samples.
    We explore some theoretical aspects of the convergence rate of our Time-varying Graph signal Reconstruction via Sobolev Smoothness (GraphTRSS) algorithm by studying the condition number of the Hessian associated with our optimization problem.
    Our algorithm has the advantage of converging faster than other methods that are based on Laplacian operators without requiring expensive eigenvalue decomposition or matrix inversions.
    The proposed GraphTRSS is evaluated on  several datasets including two COVID-19 datasets and it has outperformed many existing state-of-the-art methods for time-varying graph signal reconstruction.
    GraphTRSS has also shown excellent performance on two environmental datasets for the recovery of particulate matter and sea surface temperature signals.
\end{abstract}


% \fi
\IEEEpeerreviewmaketitle

\section{Introduction}
\label{sec:introduction}

Graphs provide the ability to model interactions of data residing on irregular and complex structures.
Social, financial, and sensor networks are examples of data that can be modeled on graphs.
Graph Signal Processing (GSP) is  an emerging field that aims to extend the concepts of classical digital signal processing to graphs \cite{sandryhaila2013discrete,shuman2013emerging,sandryhaila2014big,ortega2018graph}.
GSP has been used in several applications such as computer vision \cite{giraldo2020graph,giraldo2020graphbgs}, forecasting \cite{lewenfus2020joint}, sensor networks \cite{egilmez2014spectral}, biological networks \cite{huang2016graph}, image, and 3D point cloud processing \cite{zhang2014point,pang2015optimal,thanou2016graph}.
Similarly, the study of machine learning on graphs \cite{dong2020graph} has benefited profoundly from GSP; some examples include semi-supervised learning \cite{gadde2014active,anis2018sampling}, graph learning \cite{kalofolias2016learn,egilmez2018graph,pu2021kernel}, and the development of filters for graph convolutional networks \cite{defferrard2016convolutional,kipf2017semi}.

The sampling and recovery of graph signals are fundamental tasks in GSP that have recently received considerable attention.
Naturally, the mathematics of sampling theory and spectral graph theory have been combined, leading to generalized Nyquist sampling principles for graphs \cite{anis2014towards,shomorony2014sampling,chen2015discrete,anis2016efficient}, where most of these previous works have focused on static graph signals.
The reconstruction of time-varying  graph signals is an important problem that has not been well explored\footnote{One can think of the reconstruction of time-varying graph signals as a matrix completion problem where each row (or column) is associated with a node, and each column (or row) is associated with time.}.
The reconstruction of time-varying graph signals from discrete samples has several real world applications, such as the estimation of new cases of infectious diseases \cite{giraldo2020minimization}, or the recovery of the sea surface temperature for the study of the earth's climate dynamics \cite{qiu2017time}.

Some recent methods for the reconstruction of time-varying graph signals have also utilized the assumption that  temporal differences of graph signals are smooth.
For example, Qiu \etal \cite{qiu2017time} extended the definition of smooth signals from static to time-varying graph signals, and solved an optimization problem for reconstruction.
However, Qiu's method may have slow convergence rate because the eigenvalue spread of the Hessian associated with their optimization problem may be large.
The techniques presented in \cite{gavili2017shift,hua2020online,xia2021distributed} may be used to circumvent the problem of slow convergence.
However, these techniques require eigenvalue decomposition or matrix inversion, which are computationally expensive  for large graphs.

In the current work, we build on our previous work \cite{giraldo2020minimization}, and we propose a new algorithm to reconstruct time-varying graph signals from samples.
Our algorithm is based on the extension of the Sobolev norm defined in GSP for time-varying graph signals \cite{pesenson2009variational,giraldo2020graphbgs}.
Therefore, we dub our algorithm as Time-varying Graph signal Reconstruction via Sobolev Smoothness (GraphTRSS).
Our algorithm uses a Sobolev smoothness function to formulate an optimization problem for the reconstruction of a time-varying graphs signal from its samples, where the samples are obtained with a random sampling strategy.
The graph is constructed with a $k$-Nearest Neighbors ($k$-NN) method.
The optimization problem of GraphTRSS is solved with the conjugate gradient method.
We analyze the convergence rate of our algorithm by studying the condition number of the Hessian associated with our optimization function, and we conclude that our algorithm converges faster than its closest competitor \cite{qiu2017time} under certain conditions.
Moreover, GraphTRSS does not require expensive eigenvalue decomposition or matrix inversion.
Finally, we evaluate our algorithm on synthetic data, two COVID-19, and two environmental datasets, where our algorithm outperforms several state-of-the-art methods.

The main contributions of this work are summarized as follows:
\begin{itemize}[leftmargin=*]
    \item We use the concept of Sobolev norms from static graph signals to define a smoothness function for time-varying graph signals, and then we use this new conception to introduce an algorithm for reconstruction.
    \item We provide several mathematical insights of GraphTRSS.
    Specifically, we analyze the convergence rate of our algorithm by studying the condition number of the Hessian associated with our problem.
    \item GraphTRSS improves convergence speed without relying on expensive matrix inversion or eigenvalue decomposition.
    \item Concepts of reconstruction of graph signals from GSP are introduced in the mathematical modeling of infectious diseases for COVID-19, as well as environmental data.
\end{itemize}

The rest of the paper is organized as follows.
Section \ref{sec:related_works} shows related work in time-varying graph signals reconstruction.
Section \ref{sec:time_varying_graph_signals} presents the mathematical notations and theoretical background of this paper.
Section \ref{sec:proposed_algorithm} explains the details of GraphTRSS.
Section \ref{sec:experimental_framework} introduces the experimental framework. 
Finally, Sections \ref{sec:results_and_discussion} and \ref{sec:conclusions} show the results and conclusions, respectively.

\section{Related Work}
\label{sec:related_works}

The problems of sampling and reconstruction of graph signals have been widely explored in GSP \cite{chen2015discrete,di2016adaptive,anis2016efficient,romero2016kernel,chepuri2017graph,valsesia2018sampling,parada2019blue,venkitaraman2019predicting}.
Pesenson \cite{pesenson2008sampling} introduced concepts of Paley-Wiener spaces in graphs, where a graph signal can be determined by its samples in a set of nodes called uniqueness set.
One can say that a set of nodes is a uniqueness set of a certain graph if the fact that two graph signals in the Paley-Wiener space of the graph coincide in the uniqueness set implies that they coincide in the whole set of nodes.
As a result, a bandlimited graph signal can exactly be reconstructed from its samples if the graph signal is sampled according to its uniqueness set.
However, the bandlimitedness assumption is not realistic, \ie the graph signals in real-world datasets tend to be approximately bandlimited instead of strictly bandlimited.
Therefore, several researchers have proposed reconstruction algorithms based on the smoothness assumption of graph signals \cite{belkin2004regularization,narang2013localized,chen2016signal}, where the smoothness is measured with a Laplacian function.
Similarly, other studies have used the Total Variation of graph signals \cite{chen2015signal}, or extensions of the concept of stationarity on graph signals \cite{perraudin2017stationary,loukas2019stationary} for reconstruction.

For time-varying graph signals, scientists have developed notions of joint harmonic analysis linking together the time-domain signal processing techniques with GSP \cite{grassi2017time}, while other researchers have proposed reconstruction algorithms assuming bandlimitedness of the signals at each instant \cite{wang2015distributed,chen2015signal}.
Most of these methods do not fully exploit the underlying temporal correlations of the time-varying graph signals.
Qiu \etal \cite{qiu2017time} proposed an approach where the temporal correlations are captured with a temporal difference operator in the time-varying graph signal.
However, Qiu's method may have slow convergence because the optimization problem depends on the Laplacian matrix. In particular, the eigenvalue spread of the Hessian associated with their problem may be large, leading to poor condition numbers.
Other studies have reported the undesirable effects in convergence due to the large eigenvalue spread when dealing with shift operators derived from the Laplacian or the adjacency matrix.
For example, Hua \etal \cite{hua2020online} used a method of second-order moments to solve the problem of slow convergence speed, and this approach showed improvement in performance at the expense of additional computational cost by expensive matrix inversion operations.
We can also circumvent the problem of slow convergence by using energy-preserving shift operators as in \cite{gavili2017shift}.
For example, Xia \etal \cite{xia2021distributed} exploited an energy-preserving shift operator for distributed learning problems over graphs.
However, the energy-preserving shift operators require the eigendecomposition of the adjacency (or Laplacian) matrix, which is computational prohibitive for large graphs.

In the current work, we assume smoothness in the temporal differences of graph signals as in \cite{qiu2017time}.
However, our algorithm improves the condition number of the Hessian associated with our problem without requiring expensive eigenvalue decompositions or matrix inversions.
We show that GraphTRSS converges faster than Qiu's method \cite{qiu2017time} under some conditions.

\section{Time Varying Graph Signals}
\label{sec:time_varying_graph_signals}

\subsection{Notation}
\label{sec:notation}

In this paper, calligraphic letters such as $\mathcal{V}$ denote sets, and $\vert \mathcal{V} \vert$ represents its cardinality.
Uppercase boldface letters such as $\mathbf{W}$ represent matrices, and lowercase boldface letters such as $\mathbf{y}$ denote vectors.
$\mathbf{I}$ is the identity matrix, and $\mathbf{1}$ is a vector of ones with appropriate dimensions.
The Hadamard and Kronecker products between matrices are denoted by $\circ$ and $\otimes$, respectively.
$(\cdot)^{\mathsf{T}}$ represents transposition.
The vectorization of matrix $\mathbf{A}$ is denoted as $\vectorized{(\mathbf{A})}$, and $\diag(\mathbf{x})$ is the diagonal matrix with entries $\{x_1,x_2,\dots,x_N\}$ as its diagonal elements.
The $\ell_2$-norm of a vector is written as $\Vert \cdot \Vert_2$.
$\lambda_{\text{max}}(\mathbf{A})$ and $\lambda_{\text{min}}(\mathbf{A})$ represent the maximum and minimum eigenvalues of matrix $\mathbf{A}$, respectively.
Finally, $\kappa(\cdot)$ is the condition number and $\Vert \cdot \Vert_F$ is the Frobenius norm of a matrix.

\subsection{Preliminaries}

A graph can be represented as $G=(\mathcal{V},\mathcal{E})$, where $\mathcal{V}=\{1,2,\dots,N\}$ is the set of nodes, and $\mathcal{E}=\{(i,j)\}$ is the set of edges.
Each element in $\mathcal{E}$ represents a connection between the vertices $i$ and $j$.
The adjacency matrix $\mathbf{W}\in \mathbb{R}^{N\times N}$ represents the structure of the graph, where $\mathbf{W}(i,j) > 0 ~\forall~(i,j)\in \mathcal{E}$ represents the weight of the connection between the vertices $i$ and $j$.
In the current work, we consider $G$ as a connected, undirected, and weighted graphs.
The adjacency matrix is symmetric when we have undirected graphs, \ie the weights between $(i,j)$ and $(j,i)$ are the same.
The degree matrix $\mathbf{D}\in \mathbb{R}^{N\times N}$ is a diagonal matrix such that $\mathbf{D}=\diag(\mathbf{W1})$, \ie each element $\mathbf{D}(i,i)$ of the diagonal is the sum of all edge weights incident at the $i$th node.
In this work, we use the combinatorial Laplacian matrix defined as $\mathbf{L=D-W}$.
Since $\mathbf{L}$ is a positive semi-definite matrix, it has eigenvalues $0=\lambda_1 \leq \lambda_2 \leq \dots \leq \lambda_N$ and corresponding eigenvectors $\{ \mathbf{u}_1,\mathbf{u}_2,\dots,\mathbf{u}_N\}$.
Finally, a graph signal is a function that maps a set of nodes to the real values $x:\mathcal{V} \to \mathbb{R}$, and therefore we can represent a static graph signal as $\mathbf{x}\in \mathbb{R}^N$ where $\mathbf{x}(i)$ is the value of the graph signal at the $i$th node.

The graph Fourier operator  is defined by the eigenvalue decomposition of the Laplacian matrix $\mathbf{L} = \mathbf{U}\mathbf{\Lambda}\mathbf{U}^{\mathsf{T}}$ \cite{ortega2018graph}, where $\mathbf{U}=[\mathbf{u}_1,\mathbf{u}_2,\dots,\mathbf{u}_N]$ and $\mathbf{\Lambda}=\diag(\lambda_1,\lambda_2,\dots,\lambda_N)$.
One can associate $\lambda_i$ of $\mathbf{L}$ with the frequency of the $i$th eigenvalue \cite{ortega2018graph}.
The Graph Fourier Transform (GFT) $\mathbf{\hat{x}}$ of the graph signal $\mathbf{x}$ can be defined as $\mathbf{\hat{x}}=\mathbf{U}^{{\mathsf{T}}}\mathbf{x}$, and similarly, the inverse GFT is given by $\mathbf{x} = \mathbf{U}\mathbf{\hat{x}}$ \cite{ortega2018graph}.
A graph signal $\mathbf{x}$ is called bandlimited if we can express it with a few spectral components, \ie $\mathbf{x}$ is bandlimited if $\exists~\rho \in \{ 1,2,\dots,N-1 \}$ such that its GFT satisfies $\mathbf{\hat{x}}(i)=0~\forall~i > \rho$.
Several authors have proved that we can perfectly reconstruct a graph signal from at least $\rho$ samples if $\mathbf{x}$ is bandlimited \cite{pesenson2008sampling,chen2015discrete}.
However, real-world graph signals tend to be approximately bandlimited, and therefore researchers have focused on using approximately smooth Laplacian functions for the reconstruction of static and time-varying graph signals.

\subsection{Reconstruction of Time-Varying Graph Signals}
\label{sec:recons_time_varying_graph_signals}

The sampling and reconstruction of graph signals have played a central role in the study of GSP \cite{chen2015discrete,anis2016efficient,romero2016kernel,parada2019blue}.
Several studies have used the smoothness assumption in graphs to solve problems of reconstruction and sampling of graph signals.
Formally, notions of smoothness in $\mathbf{x}$ have been introduced with concepts of local variation \cite{shuman2013emerging}.
One can define the local variation of a graph signal $\mathbf{x}$ at node $i$ as:
\begin{equation}
    \Vert \nabla_i \mathbf{x} \Vert_2 \triangleq \left[ \sum_{j \in \mathcal{N}_i} \mathbf{W}(i,j) [\mathbf{x}(j)-\mathbf{x}(i)]^2 \right]^{\frac{1}{2}},
    \label{eqn:local_variation}
\end{equation}
where $\mathcal{N}_i$ is the set of neighbors of node $i$.
Moreover, one can introduce the \textit{discrete p-Dirichlet form} for notions of global smoothness as $S_p(\mathbf{x}) \triangleq \frac{1}{p} \sum_{i \in \mathcal{V}} \Vert \nabla_i \mathbf{x} \Vert_2^p$, then:
\begin{equation}
    S_p(\mathbf{x}) = \frac{1}{p} \sum_{i \in \mathcal{V}} \left[ \sum_{j \in \mathcal{N}_i} \mathbf{W}(i,j) [\mathbf{x}(j)-\mathbf{x}(i)]^2 \right]^{\frac{p}{2}}.
    \label{eqn:p-dirichlet_form}
\end{equation}
For example, when $p = 2$, we have $S_2(\mathbf{x})$ which is known as the graph Laplacian quadratic form \cite{shuman2013emerging}:
\begin{equation}
    S_2(\mathbf{x}) = \sum_{(i,j) \in \mathcal{E}} \mathbf{W}(i,j) [\mathbf{x}(j)-\mathbf{x}(i)]^2 = \mathbf{x}^\mathsf{T}\mathbf{Lx}.
    \label{eqn:Laplacian_quadratic_form}
\end{equation}
Notice that $S_2(\mathbf{x}) = 0 \iff \mathbf{x}=\tau \boldsymbol{1}$, with $\tau$ a constant.
The Laplacian quadratic form $S_2(\mathbf{x})$ is small when we have slight variations of the graph signal across the nodes in the graph, \ie a smooth graph signal.
Some researchers have used the graph Laplacian quadratic form as a regularization term to solve problems of reconstruction of graph signals \cite{puy2018random}.
For time-varying graph signals, an extension of the graph Laplacian quadratic form has also been used for reconstruction \cite{qiu2017time}.

Let $\mathbf{X} = [\mathbf{x}_1, \mathbf{x}_2, \dots, \mathbf{x}_M]$ be a time-varying graph signal, where $\mathbf{x}_i \in \mathbb{R}^{N}$ is a graph signal in $G$ at time $i$.
One can extend the concept of graph Laplacian quadratic form to time-varying graph signals by summing the Laplacian quadratic form of each graph signal $\mathbf{x}_i$. Then, we have:
\begin{equation}
    S_2(\mathbf{X}) = \sum_{i=1}^M \mathbf{x}_i^\mathsf{T}\mathbf{Lx}_i = \tr(\mathbf{X}^{\mathsf{T}}\mathbf{LX}).
    \label{eqn:smoothness_time_varying}
\end{equation}
However, notice that (\ref{eqn:smoothness_time_varying}) does not have temporal relationships between graph signals at different times $i$.
To include time information, Qiu \etal \cite{qiu2017time} introduced the temporal difference operator $\mathbf{D}_h \in \mathbb{R}^{M \times (M-1)}$ defined as follows:
\begin{equation}
\mathbf{D}_h = 
    \begin{bmatrix} 
    -1 &    &        &    \\
     $ $1 & -1 &        &    \\
       &  $ $1 & \ddots &    \\
       &    & \ddots & -1 \\
       &    &        &  $ $1 \\
    \end{bmatrix}
    \in \mathbb{R}^{M \times (M-1)}.
    \label{eqn:temporal_difference_operator}
\end{equation}
As a consequence, the temporal difference graph signal is such that:
\begin{equation}
    \mathbf{XD}_h = [\mathbf{x}_2-\mathbf{x}_1,\mathbf{x}_3-\mathbf{x}_2,\dots,\mathbf{x}_M-\mathbf{x}_{M-1}].
    \label{eqn:difference_signal}
\end{equation}
Qiu \etal \cite{qiu2017time} noted that $S_2(\mathbf{XD}_h)$ exhibits better smoothness properties compared to $S_2(\mathbf{X})$ in real-world datasets for time-varying graph signals, \ie the difference signal $\mathbf{x}_i-\mathbf{x}_{i-1}$ exhibits smoothness in the graph even if the signal $\mathbf{x}_i$ is not smooth across the graph. They defined the structure of time-varying graph signals as follows \cite{qiu2017time}:
\begin{definition}
    The $\alpha$-structured set $\mathcal{B}_\alpha(G)$ composed of smoothly evolving graph signals is defined as:
    \begin{equation}
        \mathcal{B}_\alpha(G) = \left\{ \mathbf{X}: \tr\left((\mathbf{XD}_h)^{\mathsf{T}}\mathbf{LXD}_h\right) \leq (M-1)\alpha  \right\},
        \label{eqn:structure_smooth_signals}
    \end{equation}
    where $\alpha$ indicates the smoothness level of the time-varying graph signal.
    \label{dfn:structure_smooth_signals}
\end{definition}

Qiu \etal \cite{qiu2017time} used Definition \ref{dfn:structure_smooth_signals} to introduce a Time-varying Graph Signal Reconstruction (TGSR) method as follows:
\begin{equation}
    \min_{\mathbf{\tilde{X}}} \frac{1}{2} \Vert \mathbf{J} \circ \mathbf{\mathbf{\tilde{X}}} - \mathbf{Y} \Vert_F^2 + \frac{\upsilon}{2} \tr\left((\mathbf{\tilde{X}D}_h)^{\mathsf{T}}\mathbf{L\tilde{X}D}_h\right),
    \label{eqn:qiu_reconstruction_noisy}
\end{equation}
where $\mathbf{J} \in \{0,1\}^{N \times M}$ is the sampling matrix of $\mathbf{X}$, ${\upsilon}$ is a regularization parameter, and $\mathbf{Y} \in \mathbb{R}^{N\times M}$ is the matrix of signals that we know (the observed values).
The sampling matrix $\mathbf{J}$ is defined as follows:
\begin{equation}
    \mathbf{J}(i,j) = \begin{cases}
        1 & \text{if } i \in \mathcal{S}_j, \\
        0 & \text{if } i \notin \mathcal{S}_j,
    \end{cases}
\end{equation}
where $\mathcal{S}_j$ is the set of sampled nodes at column $j$.
\begin{theorem}[Qiu \etal \cite{qiu2017time}]
    \label{trm:uniqueness_solution}
    The solution of (\ref{eqn:qiu_reconstruction_noisy}) is unique when the following conditions are satisfied by the sampling matrix $\mathbf{J}$:
    \begin{enumerate}
        \item For any $n \in \{1,\dots,N\},~\exists~m \in \{1,\dots,M\}$ such that $\mathbf{J}(n,m)=1$.
        \item There is a fiducial time $m_0\in \{1,\dots,M\}$, such that for any $m\in \{1,\dots, M\}$, with $m \neq m_0$, there exist a node $n_m\in \{1,\dots, N\}$ satisfying that $\mathbf{J}(n_m,m_0)=\mathbf{J}(n_m,m)=1$.
    \end{enumerate}
    Proof: see \cite{qiu2017time}.
\end{theorem}
Theorem \ref{trm:uniqueness_solution} provides some properties that $\mathbf{J}$ should satisfy to obtain a unique solution for (\ref{eqn:qiu_reconstruction_noisy}).
Notice that deterministic sampling methods \cite{chen2015discrete,anis2016efficient,tsitsvero2016signals,lorenzo2018sampling} or sampling on product graphs \cite{ortiz2018sampling} do not satisfy the first condition of Theorem \ref{trm:uniqueness_solution}, so it is not suitable for these problems.
In contrast, in the current work we propose a random sampling strategy as in \cite{qiu2017time}.

Eqn. (\ref{eqn:qiu_reconstruction_noisy}) reconstructs a time-varying graph signal $\mathbf{\tilde{X}}$ with 1) a small error $\Vert \mathbf{J} \circ \mathbf{\mathbf{\tilde{X}}} - \mathbf{Y} \Vert_F^2$, and 2) a small value of the temporal difference graph signal smoothness $\tr((\mathbf{\tilde{X}D}_h)^{\mathsf{T}}\mathbf{L\tilde{X}D}_h)$.
%Notice that the second term in (\ref{eqn:qiu_reconstruction_noisy}) enforces $(\mathbf{\tilde{X}D}_h)$ to act as eigenvectors of $\mathbf{L}$, enforcing $\mathbf{\tilde{X}}$ to have a spatiotemporal-type structure.
Finally, the parameter ${\upsilon}$ in (\ref{eqn:qiu_reconstruction_noisy}) weights the importance between the error and smoothness terms.
This parameter ${\upsilon}$ is usually tuned experimentally.

In the current work, we explore two additional temporal difference operators: a two-time steps operator and a three-time steps operator. The mathematical definitions of these operators are given in Section I of the supplementary material.

\section{Sobolev Smoothness of Time-Varying Graph Signals}
\label{sec:proposed_algorithm}

In the current work, we extend the definition of Sobolev norms in GSP \cite{pesenson2009variational,giraldo2020graphbgs} from static graph signals to a smoothness function for time-varying graph signals, and then we formulate a new reconstruction algorithm.
The Sobolev norm was defined by Pesenson \cite{pesenson2009variational} as follows:
\begin{definition}[Sobolev norm]
    \label{dfn:sobolev_norm}
    Let $\mathbf{L}$ and $\mathbf{x}$ be the Laplacian matrix and graph signal, respectively. For fixed parameters $\epsilon \geq 0$ and $\beta \in \mathbb{R}^+$, the Sobolev norm is given as follows:
    \begin{equation}
        \Vert \mathbf{x} \Vert_{\beta,\epsilon} \triangleq \Vert (\mathbf{L}+\epsilon \mathbf{I})^{\beta/2} \mathbf{x} \Vert_2.
        \label{eqn:sobolev_norm}
    \end{equation}
\end{definition}
When $\mathbf{L}$ is symmetric (as is the case in this work), we have that:
\begin{equation}
    \Vert \mathbf{x} \Vert_{\beta,\epsilon}^{2} 
    = \mathbf{x}^{\mathsf{T}}(\mathbf{L}+\epsilon\mathbf{I})^{\beta}\mathbf{x}.
    \label{eqn:sobolev_norm_rewritten}
\end{equation}
Notice the Sobolev norm in (\ref{eqn:sobolev_norm_rewritten}) is equal to the Laplacian quadratic form in (\ref{eqn:Laplacian_quadratic_form}) when $\epsilon=0$ and $\beta=1$.
We use the Sobolev norm to define a new smoothness function for time-varying graph signals as follows:
\begin{definition}[Sobolev smoothness of time-varying graph signals]
    \label{dfn:sobolev_norm_time_varying}
    Let $\mathbf{X} = [\mathbf{x}_1, \mathbf{x}_2, \dots, \mathbf{x}_M]$ be a time-varying graph signal, let $\mathbf{D}_h$ be the temporal difference operator, and let $\mathbf{L}$ be the combinatorial Laplacian matrix of a graph $G$. For fixed parameters $\epsilon \geq 0$ and $\beta \in \mathbb{R}^+$, the Sobolev smoothness of $\mathbf{X}$ is given as follows:
    \begin{gather}
        \nonumber
        S_{\beta,\epsilon}(\mathbf{X}) \triangleq \sum_{i=2}^M (\mathbf{x}_i-\mathbf{x}_{i-1})^{\mathsf{T}}(\mathbf{L}+\epsilon\mathbf{I})^{\beta}(\mathbf{x}_i-\mathbf{x}_{i-1})\\
        =\tr\left((\mathbf{XD}_h)^{\mathsf{T}}(\mathbf{L}+\epsilon\mathbf{I})^{\beta}(\mathbf{XD}_h)\right).
        \label{eqn:sobolev_norm_time}
    \end{gather}
    %\begin{equation}
    %    {\color{blue}S_{\beta,\epsilon}(\mathbf{X})} \triangleq \sum_{i=1}^M \mathbf{x}_i^{\mathsf{T}}(\mathbf{L}+\epsilon\mathbf{I})^{\beta}\mathbf{x}_i = \tr(\mathbf{X}^{\mathsf{T}}(\mathbf{L}+\epsilon\mathbf{I})^{\beta}\mathbf{X}).
    %    \label{eqn:sobolev_norm_time}
    %\end{equation}
\end{definition}

\subsection{Sobolev Reconstruction}

We use the Sobolev smoothness in Definition \ref{dfn:sobolev_norm_time_varying} to formulate two new reconstruction algorithms.
The first algorithm solves the problem for the noiseless case, while the second algorithm solves the noisy case.

In the noiseless case, we assume that the sampling mechanism does not add noise to the problem.
As a consequence, the observed graph signal is given by $\mathbf{Y} = \mathbf{J} \circ \mathbf{\tilde{X}}$.
One can formulate an optimization problem to get an approximate reconstruction of the time-varying graph signal as follows:
\begin{equation}
    \min_{\mathbf{\tilde{X}}} \frac{1}{2}\tr\left((\mathbf{\tilde{X}D}_h)^{\mathsf{T}}(\mathbf{L}+\epsilon\mathbf{I})^{\beta}\mathbf{\tilde{X}D}_h\right)~\text{s.t.}~\mathbf{J} \circ \mathbf{\tilde{X}} = \mathbf{Y}.
    \label{eqn:sobolev_min_noiseless}
\end{equation}
The optimization function in (\ref{eqn:sobolev_min_noiseless}) reconstructs a smooth spatiotemporal graph signal $\mathbf{\tilde{X}}$ given the constraint $\mathbf{Y} = \mathbf{J} \circ \mathbf{\tilde{X}}$.
The noiseless case can be solved by a gradient projection algorithm.
The iterative update is such that:
\begin{equation}
    \mathbf{\tilde{X}}^{t+1} = \left( \mathbf{\tilde{X}}^{t}-\xi \nabla_{\mathbf{\tilde{X}}} f_s(\mathbf{\tilde{X}}^{t}) \right)^+,
    \label{eqn:update_step}
\end{equation}
where $f_s(\mathbf{\tilde{X}}^{t})=\frac{1}{2}\tr\left((\mathbf{\tilde{X}}^{t}\mathbf{D}_h)^{\mathsf{T}}(\mathbf{L}+\epsilon\mathbf{I})^{\beta}\mathbf{\tilde{X}}^{t}\mathbf{D}_h\right)$, $\xi$ is the step size, $\nabla_{\mathbf{\tilde{X}}} f_s(\mathbf{\tilde{X}}^{t})$ is the gradient of function $f_s(\mathbf{\tilde{X}}^{t})$ given as:
\begin{equation}
    \nabla_{\mathbf{\tilde{X}}} f_s(\mathbf{\tilde{X}}^{t})=(\mathbf{L}+\epsilon\mathbf{I})^{\beta}\mathbf{\tilde{X}}^{t}\mathbf{D}_h\mathbf{D}_h^{\mathsf{T}},
    \label{eqn:gradient}
\end{equation}
and $\left(\mathbf{V}\right)^+$ is the projection of signal $\mathbf{V}$ to the signal space $\mathbf{Y} = \mathbf{J} \circ \mathbf{\tilde{X}}$ given as follows \cite{qiu2017time}:
\begin{equation}
    \left(\mathbf{V}\right)^+ = \mathbf{Y+V-J}\circ\mathbf{V}.
\end{equation}

\begin{figure*}
    \centering
    \includegraphics[width=\textwidth]{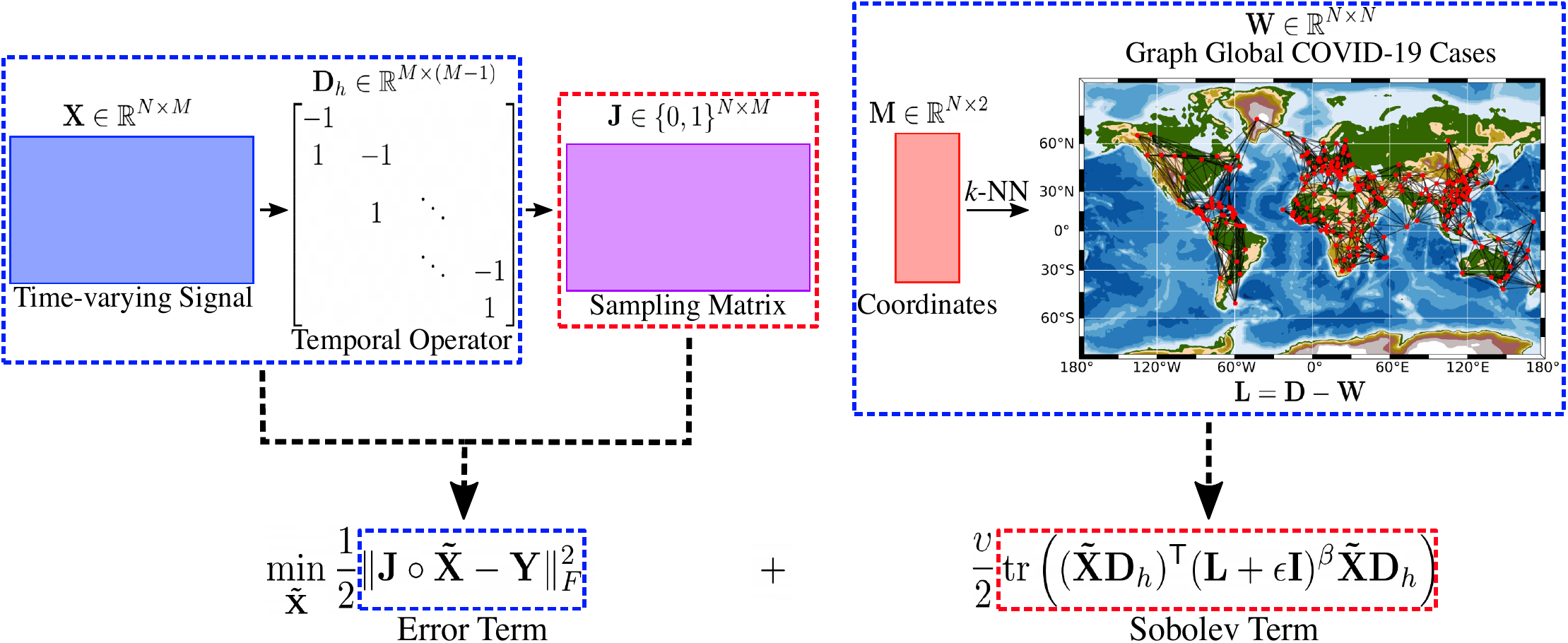}
    \caption{The framework of our algorithm (GraphTRSS) using a matrix of coordinates $\mathbf{M} \in \mathbb{R}^{N\times 2}$ to construct a graph with $N$ regions in the world with confirmed cases of COVID-19 by November 18, 2020. The graph is constructed with a $k$-Nearest Neighbors ($k$-NN) method. GraphTRSS uses the operator $\mathbf{D}_h \in \mathbb{R}^{M\times (M-1)}$ to capture temporal information in the time-varying signal $\mathbf{X} \in \mathbb{R}^{N\times M}$ with $M$ temporal snapshots, and it also uses different sampling strategies $\mathbf{J}\in \{0,1\}^{N\times M}$ according to the desired output (reconstruction or forecasting). Finally, the optimization function, which includes the error and Sobolev terms, reconstructs or predicts the missing values, \ie the indexes of $\mathbf{X}$ where $\mathbf{J}$ has values zero.}
    \label{fig:global_covid_cases}
\end{figure*}

The above formulation can be extended to the noisy case as follows.
%In this work, we focus on the noisy case because it is arguably more interesting than the noiseless case.
We take into account the noise by relaxing the constraint in (\ref{eqn:sobolev_min_noiseless}):
\begin{equation}
    \min_{\mathbf{\tilde{X}}} \frac{1}{2} \Vert \mathbf{J} \circ \mathbf{\mathbf{\tilde{X}}} - \mathbf{Y} \Vert_F^2 + \frac{{\upsilon}}{2} \tr\left((\mathbf{\tilde{X}D}_h)^{\mathsf{T}}(\mathbf{L}+\epsilon\mathbf{I})^{\beta}\mathbf{\tilde{X}D}_h\right).
    \label{eqn:sobolev_reconstruction_noisy}
\end{equation}
The optimization problem in (\ref{eqn:sobolev_reconstruction_noisy}) reconstructs a time-varying graph signal with a small error (given by the first term) and a small value for the Sobolev smoothness of its temporal difference signal.
Here, we assume that the difference of graph signals $\mathbf{XD}_h$ presents better smoothness properties than $\mathbf{X}$ alone.

In the current work, we solve \eqref{eqn:qiu_reconstruction_noisy} and \eqref{eqn:sobolev_reconstruction_noisy} using the conjugate gradient method as in \cite{qiu2017time}.
Therefore, we update the search direction $\Delta \mathbf{\tilde{X}}^t$ and the step size $\mu$ on each iteration $t$ as follows:
\begin{equation}
    \label{eqn:step_size}
    \mu = -\frac{\langle \Delta \mathbf{\tilde{X}}^t, \nabla_{\mathbf{\tilde{X}}} f_u(\mathbf{\tilde{X}}^t) \rangle}{\langle \Delta \mathbf{\tilde{X}}^t, \nabla_{\mathbf{\tilde{X}}} f_u(\Delta \mathbf{\tilde{X}}^t) + \mathbf{Y} \rangle},
\end{equation}
where $\nabla_{\mathbf{\tilde{X}}} f_u(\mathbf{\tilde{X}}) = \mathbf{J} \circ \mathbf{\tilde{X}}-\mathbf{Y} + {\upsilon} \mathbf{L}\mathbf{\tilde{X}}\mathbf{D}_h\mathbf{D}_h^{\mathsf{T}}$ to solve \eqref{eqn:qiu_reconstruction_noisy}, or $\nabla_{\mathbf{\tilde{X}}} f_u(\mathbf{\tilde{X}}) = \mathbf{J} \circ \mathbf{\tilde{X}}-\mathbf{Y} + {\upsilon} (\mathbf{L}+\epsilon \mathbf{I})^{\beta}\mathbf{\tilde{X}}\mathbf{D}_h\mathbf{D}_h^{\mathsf{T}}$ to solve \eqref{eqn:sobolev_reconstruction_noisy}.
Using $\mu$, we have that $\mathbf{\tilde{X}}^{t+1} = \mathbf{\tilde{X}}^t + \mu \Delta \mathbf{\tilde{X}}^t$, where $\Delta \mathbf{\tilde{X}}^{t} = -\nabla_{\mathbf{\tilde{X}}} f_u(\mathbf{\tilde{X}}^{t}) + \gamma \Delta \mathbf{\tilde{X}}^{t-1}$ and $\gamma=\frac{\Vert \nabla_{\mathbf{\tilde{X}}} f_u(\mathbf{\tilde{X}}^{t}) \Vert_F^2}{\Vert \nabla_{\mathbf{\tilde{X}}} f_u(\mathbf{\tilde{X}}^{t-1}) \Vert_F^2}$.
The stopping condition is given either by achieving a maximum number of iterations or when:
\begin{equation}
    \Vert \Delta \mathbf{\tilde{X}}^t \Vert_F \leq \delta,
    \label{eqn:stopping_condition}
\end{equation}
where $\delta=10^{-6}$ in the experiments for both \eqref{eqn:qiu_reconstruction_noisy} and \eqref{eqn:sobolev_reconstruction_noisy}.

Fig. \ref{fig:global_covid_cases} shows the pipeline of our algorithm applied to a graph with the regions in the world with confirmed cases of COVID-19 by November 18, 2020.
The advantage of the Sobolev minimization in (\ref{eqn:sobolev_reconstruction_noisy}) is its fast convergence with respect to the Laplacian minimization in (\ref{eqn:qiu_reconstruction_noisy}).
In the following section, we analyze some properties of the convergence rate of the optimization problems described in (\ref{eqn:qiu_reconstruction_noisy}) and (\ref{eqn:sobolev_reconstruction_noisy}).

\subsection{Convergence Rate}

Intuitively, the convergence of a gradient descent method is faster when we have well-conditioned optimization problems.
Interested readers are referred to Section II in the supplementary material to see a toy example.
Formally, the rate of convergence of a gradient descent method is at best linear.
We can accelerate this rate if we reduce the condition number of the Hessian associated with the objective function of the problem \cite{arora2017more}.

\begin{theorem}
    \label{trm:convergence_rate}
    Let $\nabla_{\mathbf{z}}^2 f_{S}(\mathbf{z})= \mathbf{Q}+[{\upsilon}(\mathbf{D}_h \mathbf{D}_h^{\mathsf{T}})\otimes (\mathbf{L}+\epsilon\mathbf{I})^{\beta}]$ and $\nabla_{\mathbf{z}}^2 f_{L}(\mathbf{z})=\mathbf{Q}+[{\upsilon}(\mathbf{D}_h\mathbf{D}_h^{\mathsf{T}})\otimes \mathbf{L}]$ be the Hessian associated with the Sobolev and Laplacian time-varying graph signal reconstruction problems in (\ref{eqn:sobolev_reconstruction_noisy}) and (\ref{eqn:qiu_reconstruction_noisy}), respectively, where $\mathbf{Q}=\diag(\vectorized{(\mathbf{J})})\in \mathbb{R}^{MN\times MN}$ and $\mathbf{J} \neq \mathbf{0}$.
    We have that:
    \begin{enumerate}
        \item If $\lambda_N,\lambda_{(D)N} \geq 1$ then $0 \leq \lambda_{min}(\nabla_{\mathbf{z}}^2 f_{S}(\mathbf{z})) \leq (\lambda_N+\epsilon)^{\beta} \lambda_{(D)N}$ and $0 \leq \lambda_{min}(\nabla_{\mathbf{z}}^2 f_{L}(\mathbf{z})) \leq \lambda_N \lambda_{(D)N}$, where $\lambda_{(D)N}=\lambda_{max}(\mathbf{D}_h \mathbf{D}_h^{\mathsf{T}})$.
        \item When $\upsilon \to \infty$, $\kappa(\nabla_{\mathbf{z}}^2 f_{S}(\mathbf{z})) \to \infty$ and $\kappa(\nabla_{\mathbf{z}}^2 f_{L}(\mathbf{z})) \to \infty$.
        \item When $\epsilon \to \infty$ and $\beta>0$, or when $\beta \to \infty$ and $\lambda_N+\epsilon > 1$, $\lambda_{max}(\nabla_{\mathbf{z}}^2 f_{S}(\mathbf{z})) \to \infty$ and so $\kappa(\nabla_{\mathbf{z}}^2 f_{S}(\mathbf{z})) \to \infty$.
    \end{enumerate}
    Proof: See Appendix \ref{app:proof_convergence}.
\end{theorem}

The main result of Theorem \ref{trm:convergence_rate} is that the upper bound of the minimum eigenvalue for the Hessian associated with GraphTRSS is looser than the corresponding upper bound of TGSR in \cite{qiu2017time} for $\epsilon > 0$, $\beta > 1$, and $\lambda_N+\epsilon \geq 1$, which favors better condition numbers for the Sobolev problem.
In practice, there is a range of values of $\epsilon$ and $\beta$ where $\kappa(\nabla_{\mathbf{z}}^2 f_{S}(\mathbf{z})) < \kappa(\nabla_{\mathbf{z}}^2 f_{L}(\mathbf{z}))$, and then the GraphTRSS converges faster than the problem described in \eqref{eqn:qiu_reconstruction_noisy}.
We compute the condition numbers of the Hessian of both GraphTRSS and TGSR in four real-world datasets in Section \ref{sec:results_and_discussion}, Fig. \ref{fig:condition_number} to show that indeed $\kappa(\nabla_{\mathbf{z}}^2 f_{S}(\mathbf{z})) < \kappa(\nabla_{\mathbf{z}}^2 f_{L}(\mathbf{z}))$ in a large range of $\epsilon$ values.
Theorem \ref{trm:convergence_rate} also shows some intuitions on how to choose the parameters for GraphTRSS and TGSR regarding convergence rate in results 2) and 3).
For example, we should keep small values of ${\upsilon}$, $\epsilon$, and $\beta$ to avoid harming the convergence rate of these algorithms.
Similarly, we should maintain small values of $\epsilon$ and $\beta$ to get benefits in the convergence rate of GraphTRSS.

%Theorem \ref{trm:convergence_rate} shows some intuitions on how to choose the parameters for the problems described in \eqref{eqn:sobolev_reconstruction_noisy} and \eqref{eqn:qiu_reconstruction_noisy} regarding convergence rate.
%For example, we should keep small values of $\lambda$, $\epsilon$, and $\beta$ to avoid harming the convergence rate of these problems.
%\textcolor{blue}{From the second result of Theorem \ref{trm:convergence_rate}, we observe that the upper bound of the minimum eigenvalue of $\nabla_{\mathbf{z}}^2 f_{S}(\mathbf{z})$ is looser than the corresponding upper bound of $\nabla_{\mathbf{z}}^2 f_{L}(\mathbf{z})$ for $\epsilon > 0$, $\beta > 1$, and $\lambda_N+\epsilon \geq 1$, which favors better condition numbers for the Sobolev problem.}
%Similarly, there is a range of values of $\epsilon$ and $\beta$ where $\kappa(\nabla_{\mathbf{z}}^2 f_{S}(\mathbf{z})) < \kappa(\nabla_{\mathbf{z}}^2 f_{L}(\mathbf{z}))$, and then the Sobolev problem converges faster than the problem described in \eqref{eqn:qiu_reconstruction_noisy}.
%In practice, we should maintain small values of $\epsilon$ and $\beta$ to get benefits in the convergence rate.

\begin{figure}
    \centering
    \includegraphics[width=0.44\textwidth]{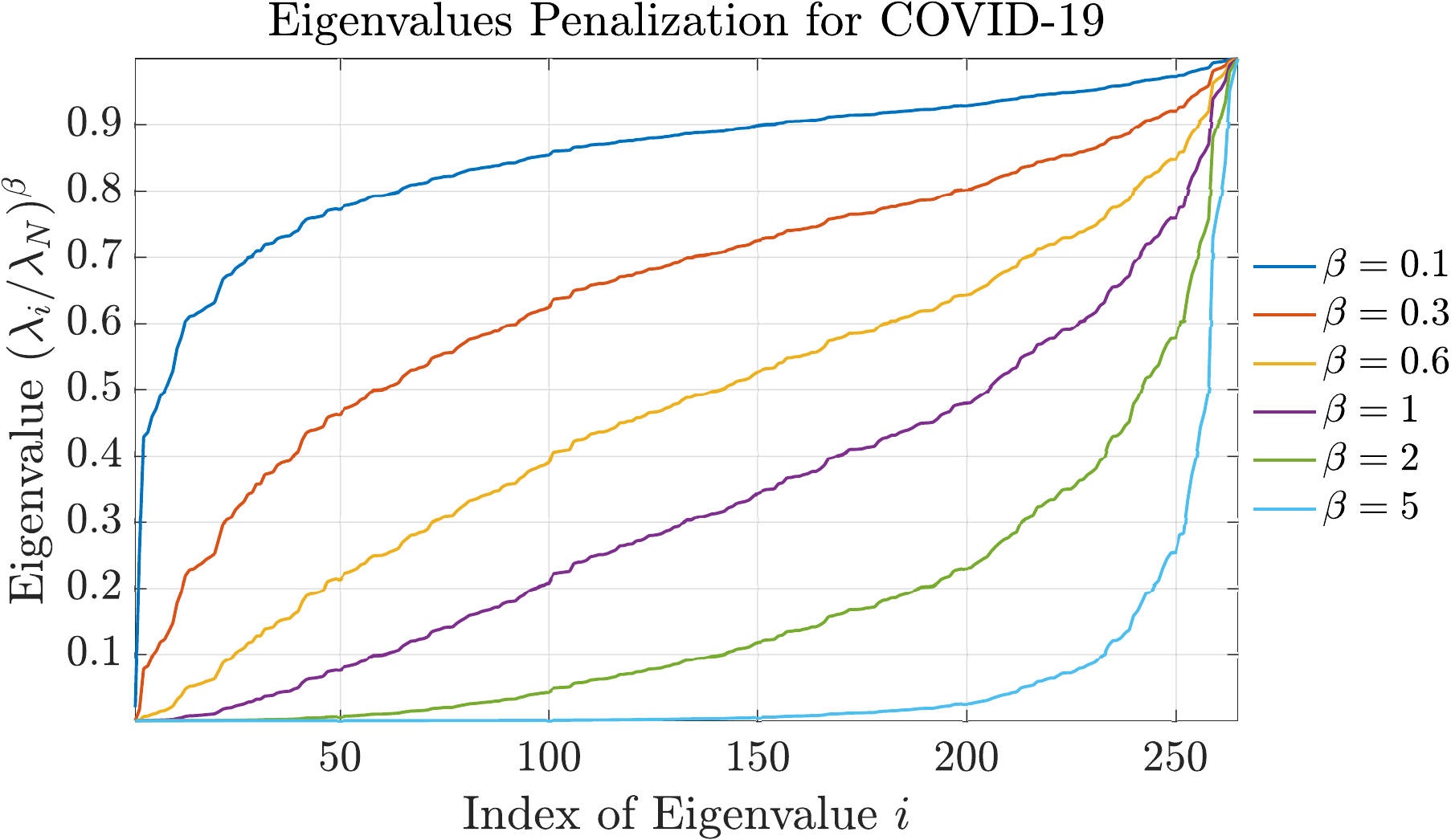}
    \caption{Eigenvalue penalization of the Laplacian matrix for different values of $\beta$ from the dataset of COVID-19.}
    \label{fig:eigenvalue_penalization}
\end{figure}

The final question is about the influence of parameter $\beta$ in the Sobolev reconstruction algorithm in (\ref{eqn:sobolev_reconstruction_noisy}).
Intuitively, the parameter $\beta$ is changing the shape of the frequencies of $\mathbf{L}$.
For simplicity, let us assume $\epsilon=0$, and let us consider only the time-varying graph signal $\mathbf{X}$ without the temporal difference operator $\mathbf{D}_h$ in (\ref{eqn:sobolev_reconstruction_noisy}).
Since the matrix of eigenvectors of $\mathbf{L}$ is an orthogonal matrix ($\mathbf{U}^{\mathsf{T}}\mathbf{U}=\mathbf{I}$), we have:
\begin{equation}
    \tr(\mathbf{X}^{\mathsf{T}}\mathbf{L}^{\beta}\mathbf{X}) = \tr(\hat{\mathbf{X}}^{\mathsf{T}}\mathbf{\Lambda}^{\beta}\hat{\mathbf{X}})=\sum_{t=1}^M \sum_{i=1}^N \hat{\mathbf{x}}_t^2(i) \lambda_i^{\beta},
    \label{eqn:penalized_laplacian}
\end{equation}
where $\hat{\mathbf{X}}=[\hat{\mathbf{x}}_1,\hat{\mathbf{x}}_2,\dots,\hat{\mathbf{x}}_M]=\mathbf{U}^{\mathsf{T}}\mathbf{X}$.
Eqn. (\ref{eqn:penalized_laplacian}) is penalizing each Fourier term of each $\hat{\mathbf{x}}_t$ with powers of the eigenvalues of $\mathbf{L}$.
For example, when $\beta=1$, we penalize the higher frequencies stronger than the lower frequencies of each $\hat{\mathbf{x}}_t$, \ie a smooth function in $G$.
For $\beta>1$, we also obtain a smooth function, but the penalization in the higher frequencies is more prominent, while for $\beta<1$, we get more penalization in the lower frequencies.
Fig. \ref{fig:eigenvalue_penalization} shows the eigenvalues of $\mathbf{L}^{\beta}$ for several values of $\beta$, where we normalized the eigenvalues such that $(\lambda_i/\lambda_N)^{\beta}~\forall~i\in\mathcal{V}$ for visualization purposes.

There are two important aspects for choosing the parameters $\epsilon$ and $\beta$ when using \eqref{eqn:sobolev_reconstruction_noisy}.
From a graph topological point of view, we notice that adding $\epsilon \mathbf{I}$ to the Laplacian adds self-loops at the vertices in the graph with weights $\epsilon$, and from Theorem \ref{trm:convergence_rate}, we notice that the value of $\epsilon$ should be small to avoid harming $\kappa(\nabla_{\mathbf{z}}^2 f_{S}(\mathbf{z}))$.
Therefore, it is reasonable to maintain a small value of $\epsilon$ so that the graph topology is not heavily modified and also to avoid harming $\kappa(\nabla_{\mathbf{z}}^2 f_{S}(\mathbf{z}))$.
Secondly, we also know from Theorem \ref{trm:convergence_rate} that choosing a large value of $\beta$ harms $\kappa(\nabla_{\mathbf{z}}^2 f_{S}(\mathbf{z}))$.
In practice, different datasets can benefit from specific penalization shapes like the ones in Fig. \ref{fig:eigenvalue_penalization}.
As a result, we set $\beta=1$ in the experiments to maintain a good $\kappa(\nabla_{\mathbf{z}}^2 f_{S}(\mathbf{z}))$ and be as fair as possible on the comparison with Qiu's method \cite{qiu2017time}.
However, we do additional experiments where we see that different values of $\beta$ can improve the performance of \eqref{eqn:sobolev_reconstruction_noisy} in different datasets, \ie different datasets require different assumptions.
Finally, we can conclude that TGSR by Qiu \etal \cite{qiu2017time} is a specific case of GraphTRSS when $\epsilon=0$ and $\beta=1$.

\section{Experimental Framework}
\label{sec:experimental_framework}

This section presents the datasets used in this work and the experimental framework details.
We divide our experiments into three parts: 1) synthetic dataset, 2) COVID-19 datasets, and 3) environmental datasets.
The graph $G$ can be constructed based on the coordinate locations of the nodes in each dataset with  $k$-NN algorithm.
Let $\mathbf{M} \in \mathbb{R}^{N\times 2}$ be the matrix of coordinates of all nodes such that $\mathbf{M} = [\mathbf{m}_1,\dots,\mathbf{m}_N]^{\mathsf{T}}$, where $\mathbf{m}_i \in \mathbb{R}^2$ is the vector with the latitude and the longitude of node $i$.
The weights of each edge $(i,j)$ are given by the Gaussian kernel:
\begin{equation}
    \mathbf{W}(i,j) = \exp{\left({-\frac{\Vert \mathbf{m}_i - \mathbf{m}_j \Vert_2^2}{\sigma^2}}\right)},
    \label{eqn:gaussian_kernel}
\end{equation}
where $\Vert \mathbf{m}_i - \mathbf{m}_j \Vert_2^2$ is the Euclidean distance between the vertices $i$ and $j$, and $\sigma$ is the standard deviation given by $\sigma = \frac{1}{\vert \mathcal{E} \vert}\sum_{(i,j) \in \mathcal{E}} \Vert \mathbf{m}_i - \mathbf{m}_j \Vert_2$.
The Gaussian kernel in (\ref{eqn:gaussian_kernel}) assigns higher weights to geographically-close locations and vice versa.
The construction of the graph depends on the specific application, and GraphTRSS is not sensitive to the graph construction methods.

\subsection{Datasets}

Our algorithm is evaluated on one synthetic dataset and four real-world datasets including, 1) a synthetic graph, 2) global COVID-19 cases \cite{dong2020interactive}, 3) USA COVID-19 cases \cite{dong2020interactive}, 4) mean concentration of Particulate Matter (PM) 2.5 \cite{qiu2017time}, and 5) sea surface temperature \cite{qiu2017time}.

\subsubsection{Synthetic Graph and Signals}

We use the synthetic dataset created by Qiu \etal \cite{qiu2017time}, where $100$ nodes are generated randomly from the uniform distribution in a $100 \times 100$ square area.
Therefore, Qiu \etal \cite{qiu2017time} used a $k$-NN to construct the graph.
The time-varying graph signal is generated with the recursive function $\mathbf{x}_t = \mathbf{x}_{t-1}+\mathbf{L}^{-\frac{1}{2}}\mathbf{f}_t$, where: 1) $\mathbf{x}_{1}$ is a low-frequency graph signal with energy $10^4$, 2) $\mathbf{L}^{-\frac{1}{2}} = \mathbf{U}\mathbf{\Lambda}^{-\frac{1}{2}}\mathbf{U}^{\mathsf{T}}$, where $\mathbf{\Lambda}^{-\frac{1}{2}}=\diag(0,\lambda_2^{-\frac{1}{2}},\dots,\lambda_N^{-\frac{1}{2}})$, and 3) $\mathbf{f}_t$ is an i.i.d. white Gaussian signal such that $\Vert \mathbf{f}_t \Vert_{2} = \alpha$.
As a result, the synthetic time-varying graph signal satisfies Definition \ref{dfn:structure_smooth_signals}.

\subsubsection{Global COVID-19}

We use the global COVID-19 dataset provided by the Johns Hopkins University \cite{dong2020interactive}.
This dataset contains the cumulative number of daily COVID-19 cases for 265 locations in the world (Fig. \ref{fig:global_covid_cases} shows the locations in the dataset), and we use the data between January 22, 2020, and November 18, 2020; \ie $302$ days.
The graph is constructed with the $k$-NN method with $k=10$ using the coordinates of each place \cite{ortega2018graph}.

%(the locations with coordinates $\{0,0\}$ were deleted from our experiments because these localities do not contain spatial information, \eg the Diamond Princess cruise).

\begin{figure}
    \centering
    \includegraphics[width=0.44\textwidth]{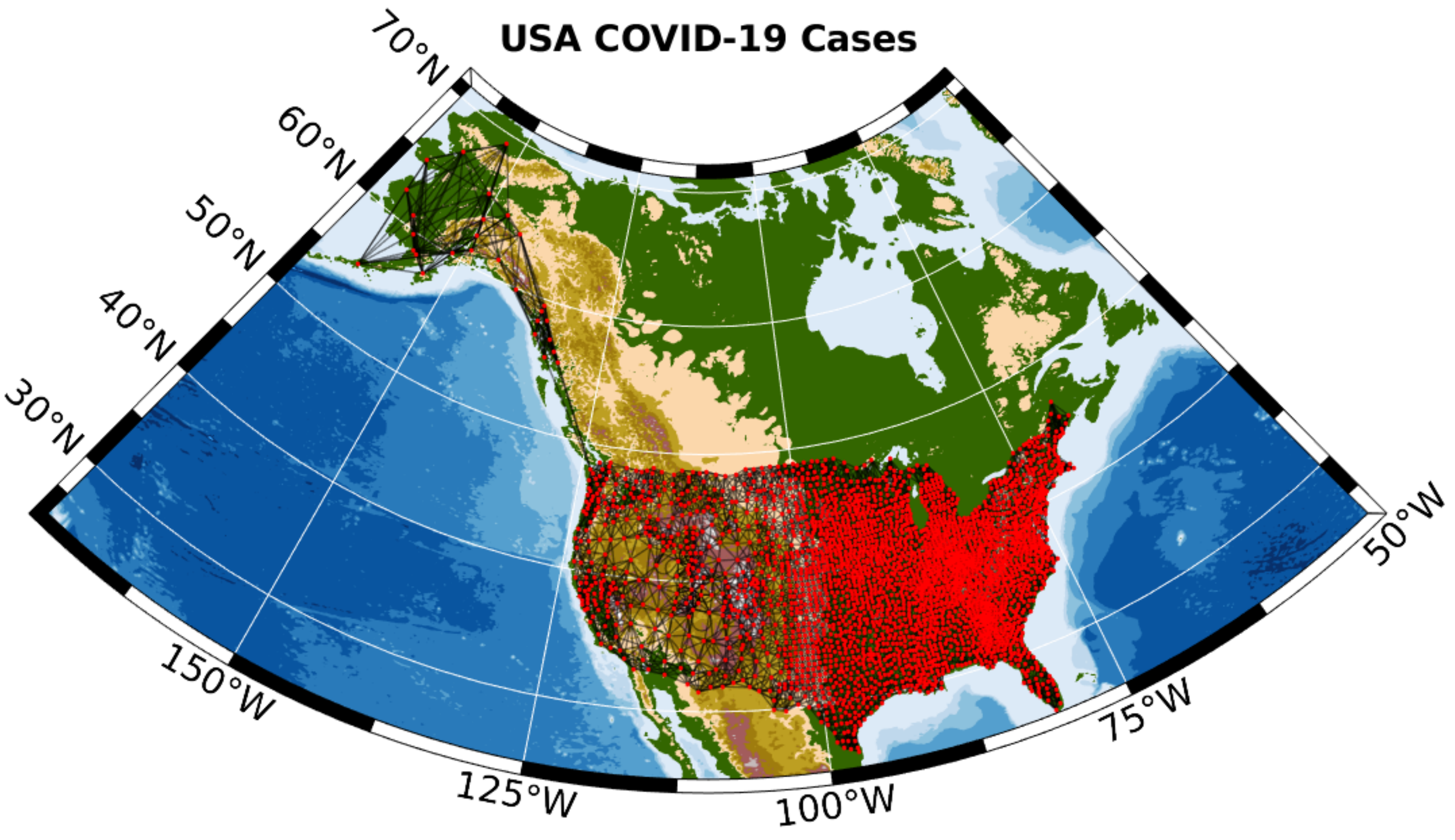}
    \caption{Graph with the places in the United States in the Johns Hopkins University dataset \cite{dong2020interactive}. The graph was constructed with a $k$-NN with $k=10$.}
    \label{fig:usa_map}
\end{figure}

\subsubsection{USA COVID-19}

We also use the USA COVID-19 dataset provided by the Johns Hopkins University \cite{dong2020interactive}.
Fig. \ref{fig:usa_map} shows the map with the points of the graph in the USA COVID-19 dataset.
This dataset is more fine-grained than the global set since the data contain $3232$ localities in the US.
We use the same temporal window and $k$-NN method as in the global dataset.
The experiments related to the US can give us a different view of our algorithm since several travel restrictions were applied between countries in the pandemic, while inside countries, there was slightly more freedom to move around.

\subsubsection{Particulate  Matter 2.5}

In this work, we use the daily mean PM 2.5 concentration dataset from California provided by the US Environmental Protection Agency\footnote{\url{https://www.epa.gov/outdoor-air-quality-data}}.
We use the data captured daily from 93 sensors in California for the first 220 days in 2015.

\subsubsection{Sea Surface Temperature}

\begin{figure}
    \centering
    \includegraphics[width=0.38\textwidth]{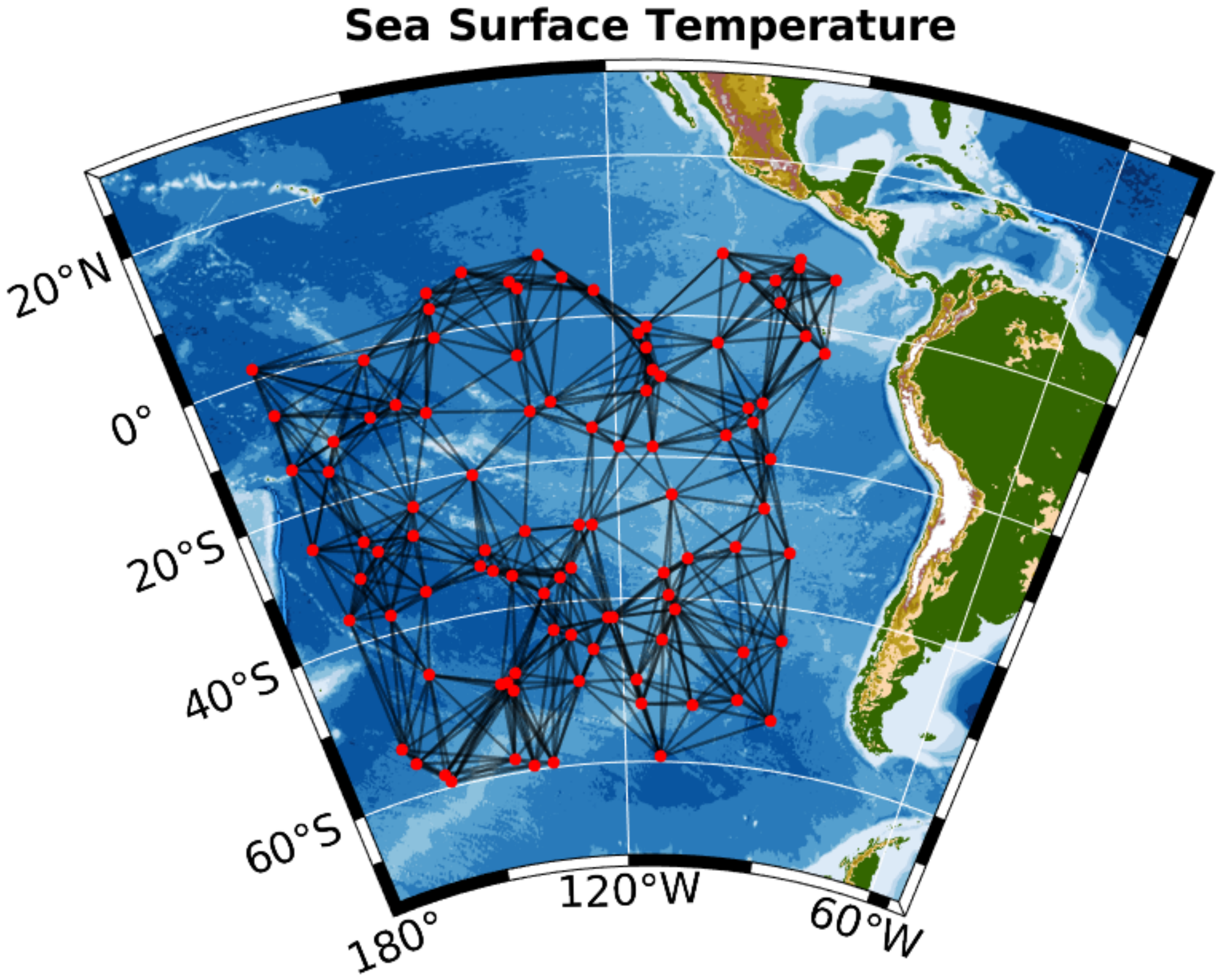}
    \caption{Graph with the spots in the sea for the dataset of temperature. The graph was constructed with a $k$-NN with $k=10$.}
    \label{fig:sea_surface_temperature}
\end{figure}

We use the sea surface temperature captured monthly and provided by the NOAA Physical Sciences Laboratory (PSL) from their website\footnote{\url{https://psl.noaa.gov/}}.
In this work, we use the same experimental framework as in \cite{qiu2017time} for comparison purposes, \ie we use a subset of 100 points on the Pacific Ocean within a time frame of 600 months.
Fig. \ref{fig:sea_surface_temperature} shows the locations of the spots in the sea for this dataset.

\subsection{Evaluation Metrics}

We use the Root Mean Square Error (RMSE), Mean Absolute Error (MAE), and the Mean Absolute Percentage Error (MAPE) to compare GraphTRSS with the methods of the literature.
$\text{RMSE}=\sqrt{\frac{\sum_{i = 1}^{N_x} ( \hat{\mathbf{x}}_i - \mathbf{x}^*_i )^2}{N_x}}$, $\text{MAE}=\frac{\sum_{i = 1}^{N_x} \vert \hat{\mathbf{x}}_i - \mathbf{x}^*_i \vert}{N_x}$, and $\text{MAPE}=\frac{1}{N_x} \sum_{i = 1}^{N_x} \vert \frac{\mathbf{x}^*_i-\hat{\mathbf{x}}_i}{\mathbf{x}^*_i} \vert$, where $\hat{\mathbf{x}}$ is the recovered signal, $\mathbf{x}^*$ is the ground truth signal, and $N_x$ is the length of the signal.

\subsection{Experiments}

For the synthetic dataset, we perform experiments analyzing several sampling densities, Signal-to-Noise Ratio (SNR) levels, smoothness $\alpha$ of the synthetic time-varying graph signal, and the condition number of $\nabla_{\mathbf{z}}^2 f_{S}(\mathbf{z})$ and $\nabla_{\mathbf{z}}^2 f_{L}(\mathbf{z})$.
For the real dataset, we perform several experiments for reconstruction and forecasting.
In the same way, we study the convergence, the $\epsilon$ and $\beta$ parameters, and the temporal difference operators for GraphTRSS.
We compare GraphTRSS with Natural Neighbor Interpolation (NNI) by Sibson \cite{sibson1981brief}, Graph Regularization (GR) by Narang \etal \cite{narang2013localized}, Tikhonov regularization by Perraudin \etal \cite{perraudin2017towards,loukas2019stationary}, Time-varying Graph Signal Reconstruction (TGSR) by Qiu \etal \cite{qiu2017time}, and Random Sampling and Decoder (RSD) by Puy \etal \cite{puy2018random}.
NNI \cite{sibson1981brief} is based on Voronoi tessellation, where the interpolation is given by a linear combination of the neighbor points of the element to be interpolated in the Voronoi partition.
GR \cite{narang2013localized} can be viewed as the solution of \eqref{eqn:qiu_reconstruction_noisy} without the time component.
Tikhonov regularization \cite{perraudin2017towards} is based on joint stationarity, where the solution is constrained to be smooth in the graph and in time.
RSD method \cite{puy2018random} solves an optimization problem where we have smoothness in the graph and a probability distribution function for the sampling.
The role of RSD and GR in the experiments is to see how these static graph signal methods compare to time-varying graph signal algorithms.
Similarly, since RSD and GR cannot reconstruct graph signals where we do not have any sample at a specific time $t$, they are not included in some of the experiments as in forecasting.
GR, Tikhonov, TGSR, and GraphTRSS have a maximum number of $20000$ iterations in the optimization algorithm for a fair comparison.
We optimize the parameters of each method in all experiments for a fair comparison.

In the reconstruction experiments, we adopt two sampling strategies.
The first reconstruction experiment performs a random sampling on each temporal snapshots for the number of nodes $N$, \ie we randomly select a percentage of nodes on each time step as in \cite{qiu2017time}.
For RSD by Puy \etal \cite{puy2018random}, we use their optimal sampling procedure because their reconstruction algorithm depends on their specific sampling distribution.
The interested readers are referred to Section III in the supplementary material for further details about RSD.
The second reconstruction experiment performs a random sampling of entire time-snapshots for the total number of graph signals $M$.
Both experiments compute the error metrics for each method on the non-sampled nodes for a set of sampling densities $\mathcal{M}$.
The set $\mathcal{M}$ was chosen according to each dataset, for example, $\mathcal{M}=\{0.5, 0.6, \dots, 0.9, 0.995\}$ for COVID-19 datasets.
Section IV in the supplementary material shows a small analysis of the bandwidths of $\mathbf{XD}_h$ for COVID-19 global.
We evaluate each method with a Monte Carlo cross-validation with $100$ repetitions for each sampling density ($50$ repetitions for the COVID-19 USA dataset).
Deterministic sampling methods do not work well with these reconstruction methods as stated in Section \ref{sec:recons_time_varying_graph_signals}.
However, the interested readers are referred to Section V in the supplementary material for a small experiment with the deterministic sampling methods in \cite{chen2015discrete,anis2018sampling,tsitsvero2016signals}.

In the forecasting experiment, we compute the error metrics for several temporal snapshots $t$ in the set $\{1,2,\dots,10\}$.
For example, since COVID-19 datasets are sampled daily, we will predict the new COVID-19 cases in the last day, in the two last days, and so on until the last ten days.
However, for the sea surface temperature, we will predict up to ten months since this dataset is sampled monthly, \ie we are interpolating multiple time steps ahead in all datasets.

We perform some studies to analyze the parameters of GraphTRSS.
The first of these studies computes the average RMSE (in $\mathcal{M}$) and the average number of iterations required to satisfy the stopping condition with variations in $\epsilon$.
The second study performs reconstruction with the three temporal difference operators (one, two, and three steps).
The last study computes reconstruction with several values of $\beta$.
These studies are evaluated with a Monte Carlo cross-validation with ten repetitions.

\begin{figure*}[h]
    \centering
    \includegraphics[width=\textwidth]{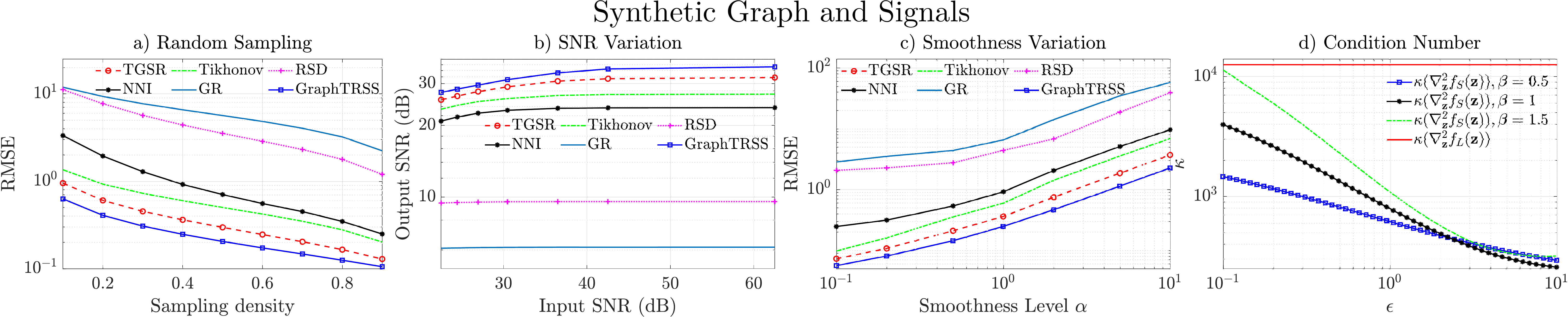}
    \caption{Comparison of GraphTRSS with several methods in the literature on synthetic data for four experiments on: a) reconstruction with several sampling densities, b) variation of the SNR, c) variation of the smoothness level in \eqref{eqn:structure_smooth_signals}, and d) $\kappa(\nabla_{\mathbf{z}}^2 f_{S}(\mathbf{z}))$ and $\kappa(\nabla_{\mathbf{z}}^2 f_{L}(\mathbf{z}))$ with several values of $\epsilon$ and $\beta$. Our algorithm is compared with Natural Neighbor Interpolation (NNI) \cite{sibson1981brief}, Graph Regularization (GR) \cite{narang2013localized}, Tikhonov regularization \cite{perraudin2017towards,loukas2019stationary}, Time-varying Graph Signal Reconstruction (TGSR) \cite{qiu2017time}, and Random Sampling and Decoder (RSD) \cite{puy2018random}.}
    \label{fig:results_synthetic_data}
\end{figure*}

\begin{table*}[]
\centering
\caption{Summary of the average error metrics in the real datasets for several sampling schemes. The best and second-best performing methods on each category are shown in {\color{red}\textbf{red}} and {\color{blue}\textbf{blue}}, respectively.}
\label{tbl:summary_results}
\makebox[\linewidth]{
\scalebox{0.97}{
\begin{tabular}{l|ccc|ccc|ccc|ccc}
\hline
\multirow{3}{*}{Method} & \multicolumn{12}{c}{Random Sampling} \\ \cline{2-13}
 & \multicolumn{3}{c|}{COVID-19 Global} & \multicolumn{3}{c|}{COVID-19 USA} & \multicolumn{3}{c|}{Sea Surface Temperature} & \multicolumn{3}{c}{PM 2.5 Concentration} \\
 & RMSE & MAE & MAPE & RMSE & MAE & MAPE & RMSE & MAE & MAPE & RMSE & MAE & MAPE \\
\hline
\scriptsize NNI (Sibson \cite{sibson1981brief}) & $1555.14$ & $189.15$ & \color{blue}$\textbf{4.23}$ & $54.92$ & $10.90$ & $2.17$ & \color{blue}$\textbf{0.77}$ & \color{blue}$\textbf{0.56}$ & \color{blue}$\textbf{0.07}$ & $4.95$ & $2.96$ & $0.59$ \\
\scriptsize GR (Narang \etal \cite{narang2013localized}) & $4744.25$ & $1119.07$ & $90.95$ & $65.49$ & $13.94$ & $3.47$ & $2.43$ & $1.76$ & $0.40$ & $5.37$ & $3.32$ & $0.67$ \\
\scriptsize Tikhonov (Perraudin \etal \cite{perraudin2017towards}) & $1253.71$ & \color{blue}$\textbf{183.60}$ & $6.07$ & $37.00$ & \color{blue}$\textbf{6.03}$ & \color{red}$\textbf{1.03}$ & $0.95$ & $0.70$ & $0.12$ & $4.32$ & $2.66$ & $0.55$ \\
\scriptsize TGSR (Qiu \etal \cite{qiu2017time}) & \color{blue}$\textbf{1136.39}$ & $213.96$ & $11.30$ & \color{blue}$\textbf{33.51}$ & $6.14$ & $1.19$ & \color{red}$\textbf{0.36}$ & \color{red}$\textbf{0.26}$ & \color{red}$\textbf{0.03}$ & \color{blue}$\textbf{3.90}$ & \color{blue}$\textbf{2.28}$ & \color{blue}$\textbf{0.39}$ \\
\scriptsize RSD (Puy \etal \cite{puy2018random}) & $2045.37$ & $506.30$ & $36.17$ & $58.59$ & $12.05$ & $2.50$ & $5.56$ & $4.62$ & $0.97$ & $5.43$ & $3.46$ & $0.72$ \\
\hline
GraphTRSS (ours) & \color{red}$\textbf{1134.15}$ & \color{red}$\textbf{152.76}$ & \color{red}$\textbf{2.41}$ & \color{red}$\textbf{33.47}$ & \color{red}$\textbf{5.96}$ & \color{blue}$\textbf{1.10}$ & \color{red}$\textbf{0.36}$ & \color{red}$\textbf{0.26}$ & \color{red}$\textbf{0.03}$ & \color{red}$\textbf{3.83}$ & \color{red}$\textbf{2.21}$ & \color{red}$\textbf{0.38}$ \\
\hline
\hline
\multirow{3}{*}{Method} & \multicolumn{12}{c}{Entire Snapshots Sampling} \\ \cline{2-13}
 & \multicolumn{3}{c|}{COVID-19 Global} & \multicolumn{3}{c|}{COVID-19 USA} & \multicolumn{3}{c|}{Sea Surface Temperature} & \multicolumn{3}{c}{PM 2.5 Concentration} \\
 & RMSE       & MAE       & MAPE       & RMSE      & MAE      & MAPE      & RMSE          & MAE          & MAPE         & RMSE         & MAE         & MAPE        \\
\hline
\scriptsize NNI (Sibson \cite{sibson1981brief}) & $1342.08$ & \color{blue}$\textbf{170.64}$ & \color{blue}$\textbf{3.40}$ & $56.69$ & $11.71$ & $2.60$ & \color{blue}$\textbf{0.98}$ & \color{blue}$\textbf{0.68}$ & \color{blue}$\textbf{0.08}$ & $10.87$ & $4.44$ & $0.85$ \\
\scriptsize Tikhonov (Perraudin \etal \cite{perraudin2017towards}) & \color{blue}$\textbf{1255.05}$ & $182.97$ & $5.94$ & $36.96$ & \color{blue}$\textbf{6.12}$ & \color{blue}$\textbf{1.04}$ & $1.17$ & $0.86$ & $0.13$ & \color{blue}$\textbf{4.54}$ & \color{blue}$\textbf{2.81}$ & \color{blue}$\textbf{0.56}$ \\
\scriptsize TGSR (Qiu \etal \cite{qiu2017time}) & $1459.45$ & $788.10$ & $100.06$ & \color{blue}$\textbf{36.38}$ & $13.75$ & $4.74$ & $19.19$ & $19.15$ & $2.13$ & $10.94$ & $9.31$ & $1.49$ \\
\hline
GraphTRSS (ours) & \color{red}$\textbf{1114.16}$ & \color{red}$\textbf{143.69}$ & \color{red}$\textbf{1.49}$ & \color{red}$\textbf{32.96}$ & \color{red}$\textbf{5.45}$ & \color{red}$\textbf{0.88}$ & \color{red}$\textbf{0.94}$ & \color{red}$\textbf{0.67}$ & \color{red}$\textbf{0.07}$ & \color{red}$\textbf{4.42}$ & \color{red}$\textbf{2.70}$ & \color{red}$\textbf{0.48}$ \\
\hline
\hline
\multirow{3}{*}{Method} & \multicolumn{12}{c}{Forecasting} \\ \cline{2-13}
 & \multicolumn{3}{c|}{COVID-19 Global} & \multicolumn{3}{c|}{COVID-19 USA} & \multicolumn{3}{c|}{Sea Surface Temperature} & \multicolumn{3}{c}{PM 2.5 Concentration} \\
 & RMSE       & MAE       & MAPE       & RMSE      & MAE      & MAPE      & RMSE          & MAE          & MAPE         & RMSE         & MAE         & MAPE        \\
\hline
\scriptsize NNI (Sibson \cite{sibson1981brief}) & $16722.30$ & $2371.30$ & \color{blue}$\textbf{6.43}$ & $904.23$ & $283.23$ & $19.54$ & $3.04$ & $2.23$ & \color{blue}$\textbf{0.23}$ & $30.71$ & $17.15$ & $2.76$ \\
\scriptsize Tikhonov (Perraudin \etal \cite{perraudin2017towards}) & $3612.06$ & \color{blue}$\textbf{1190.20}$ & $30.48$ & $99.88$ & \color{blue}$\textbf{31.2}$ & \color{blue}$\textbf{2.03}$ & \color{blue}$\textbf{1.94}$ & \color{blue}$\textbf{1.53}$ & $0.30$ & \color{blue}$\textbf{4.80}$ & \color{blue}$\textbf{3.23}$ & \color{blue}$\textbf{0.67}$ \\
\scriptsize TGSR (Qiu \etal \cite{qiu2017time}) & \color{blue}$\textbf{3313.50}$ & $2472.08$ & $238.65$ & \color{blue}$\textbf{91.5}$ & $54.36$ & $6.88$ & $18.72$ & $18.69$ & $2.39$ & $10.09$ & $9.11$ & $1.27$ \\
\hline
GraphTRSS (ours) & \color{red}$\textbf{2416.30}$ & \color{red}$\textbf{583.36}$ & \color{red}$\textbf{4.80}$ & \color{red}$\textbf{75.33}$ & \color{red}$\textbf{20.65}$ & \color{red}$\textbf{0.86}$ & \color{red}$\textbf{1.20}$ & \color{red}$\textbf{0.98}$ & \color{red}$\textbf{0.13}$ & \color{red}$\textbf{4.36}$ & \color{red}$\textbf{2.83}$ & \color{red}$\textbf{0.54}$ \\
\hline
\end{tabular}
}
}
\end{table*}

We also compare the loss function vs the iteration number for TGSR and GraphTRSS in reconstruction, where the loss is the evaluation of \eqref{eqn:qiu_reconstruction_noisy} and \eqref{eqn:sobolev_reconstruction_noisy} at each iteration.
In this case, $50$ repetitions for each sampling densities in $\mathcal{M}$ are performed, where each repetition is computed with the best parameters of each method.
For a fair comparison, GraphTRSS and Qiu's method use the same sampling matrix $\mathbf{J}$.
We also compute the running times for several sampling densities.
Finally, we compute $\kappa(\nabla_{\mathbf{z}}^2 f_{S}(\mathbf{z}))$ and $\kappa(\nabla_{\mathbf{z}}^2 f_{L}(\mathbf{z}))$ for $\beta=1$ and several values of $\epsilon$ in all real datasets.
All experiments were executed in MATLAB R2017b on a 2.3GHz MacBook Pro with 16GB memory.
The code of GraphTRSS has been made available\footnote{\url{https://github.com/jhonygiraldo/GraphTRSS}}.

\section{Results and Discussion}
\label{sec:results_and_discussion}

\begin{figure*}
    \centering
    \includegraphics[width=\textwidth]{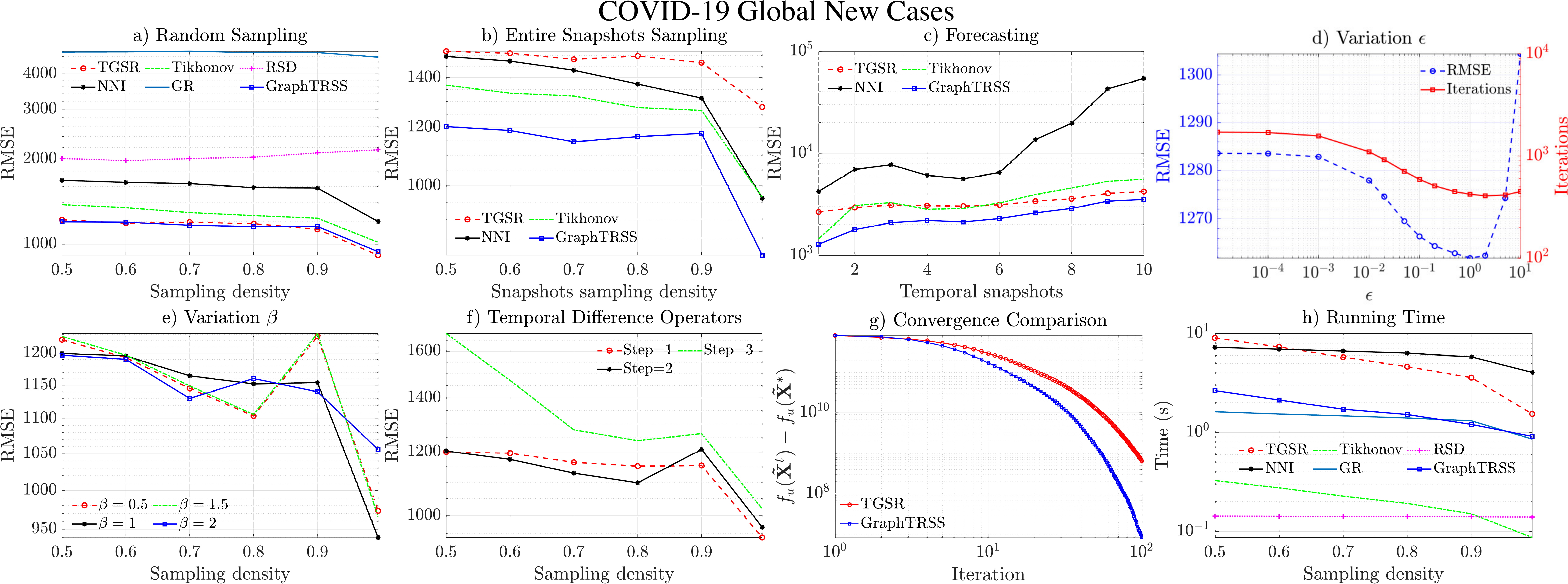}
    \caption{Comparison of GraphTRSS with several methods in the literature on COVID-19 global dataset for several experiments in terms of: a) random sampling, b) entire snapshots sampling, c) forecasting, d) variation of parameter $\epsilon$, e) variation of parameter $\beta$, f) several temporal difference operators, g) convergence comparison, and h) running time.}
    \label{fig:results_covid_global}
\end{figure*}

\begin{figure*}[h]
    \centering
    \includegraphics[width=0.99\textwidth]{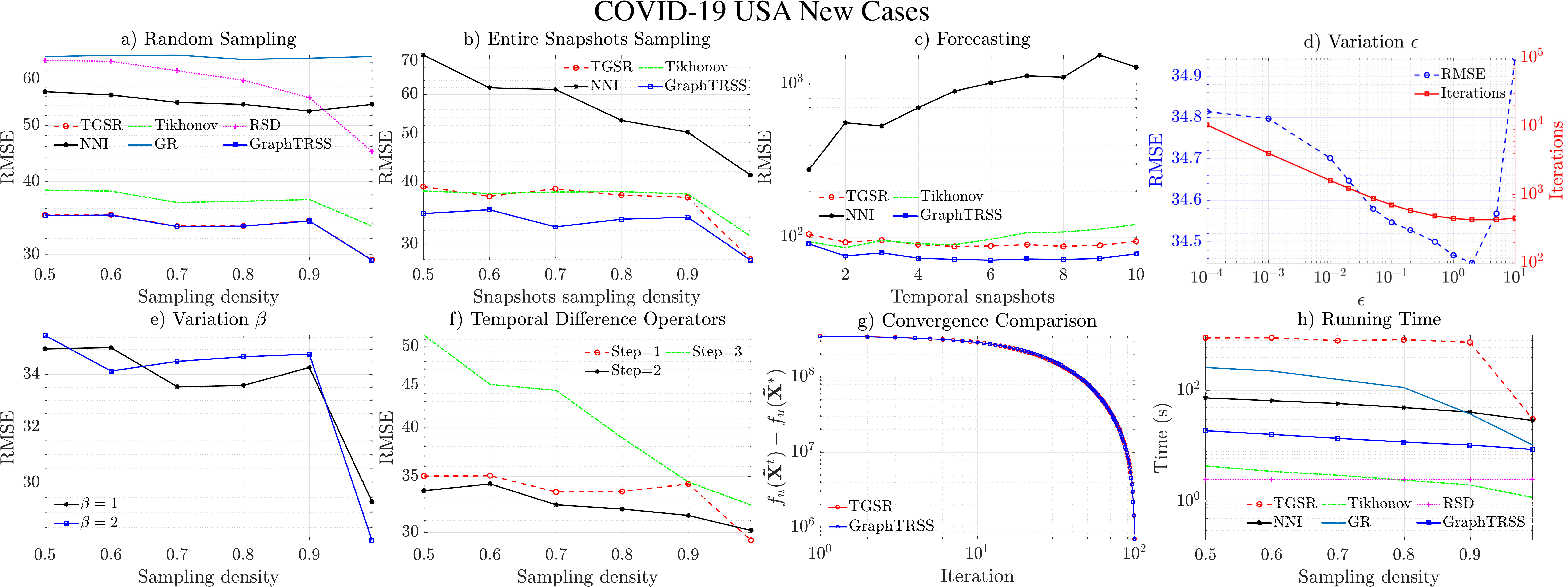}
    \caption{Comparison of GraphTRSS with several methods in the literature on COVID-19 USA dataset for several experiments in terms of: a) random sampling, b) entire snapshots sampling, c) forecasting, d) variation of parameter $\epsilon$, e) variation of parameter $\beta$, f) several temporal difference operators, g) convergence comparison, and h) running time.}
    \label{fig:results_covid_USA}
\end{figure*}

This section presents the results of the experiments on the selected datasets.
GR and RSD methods cannot reconstruct entire temporal snapshots because these algorithms only use spatial information.
Therefore, GR and RSD are not included in the experiments of entire snapshots sampling and forecasting.
A detailed analysis for each dataset is provided as follows.

\subsection{Synthetic Graph and Signals}

Fig. \ref{fig:results_synthetic_data} shows the results in the synthetic dataset.
GraphTRSS performs better than the existing methods for all values in $\mathcal{M}$ for the random sampling scheme, as shown in Fig. \ref{fig:results_synthetic_data}(a).
Our algorithm is also more robust against noise than the compared methods, as shown in Fig. \ref{fig:results_synthetic_data}(b).
Similarly, Fig. \ref{fig:results_synthetic_data}(c) shows the benefits of GraphTRSS regarding different levels of smoothness $\alpha$ as given in Definition \ref{dfn:structure_smooth_signals}.
Our algorithm shows better performance than other time-varying graph signal methods as TGSR and other static graph signal reconstruction schemes as GR.
Finally, Fig. \ref{fig:results_synthetic_data}(d) shows the condition numbers of the Hessian associated with GraphTRSS ($\kappa(\nabla_{\mathbf{z}}^2 f_{S}(\mathbf{z}))$) and TGSR ($\kappa(\nabla_{\mathbf{z}}^2 f_{L}(\mathbf{z}))$) with several values of $\epsilon$ and $\beta$.
As noted from Theorem \ref{trm:convergence_rate}, we have that: 1) there is a range of values of $\epsilon>0$ where $\kappa(\nabla_{\mathbf{z}}^2 f_{S}(\mathbf{z})) < \kappa(\nabla_{\mathbf{z}}^2 f_{L}(\mathbf{z}))$, 2) $\epsilon>0$ promotes good values of the condition number $\kappa(\nabla_{\mathbf{z}}^2 f_{S}(\mathbf{z}))$ for GraphTRSS, and 3) big values of $\beta$ harm the condition number $\kappa(\nabla_{\mathbf{z}}^2 f_{S}(\mathbf{z}))$ of GraphTRSS.

\subsection{Real Datasets Summary}

Table \ref{tbl:summary_results} shows the summary of all error metrics for random, entire snapshots sampling, and forecasting in the real datasets.
The results are computed as the mean of all results over the set $\mathcal{M}$.
GraphTRSS shows the best performance in almost all cases against the other methods.

\begin{figure*}
    \centering
    \includegraphics[width=\textwidth]{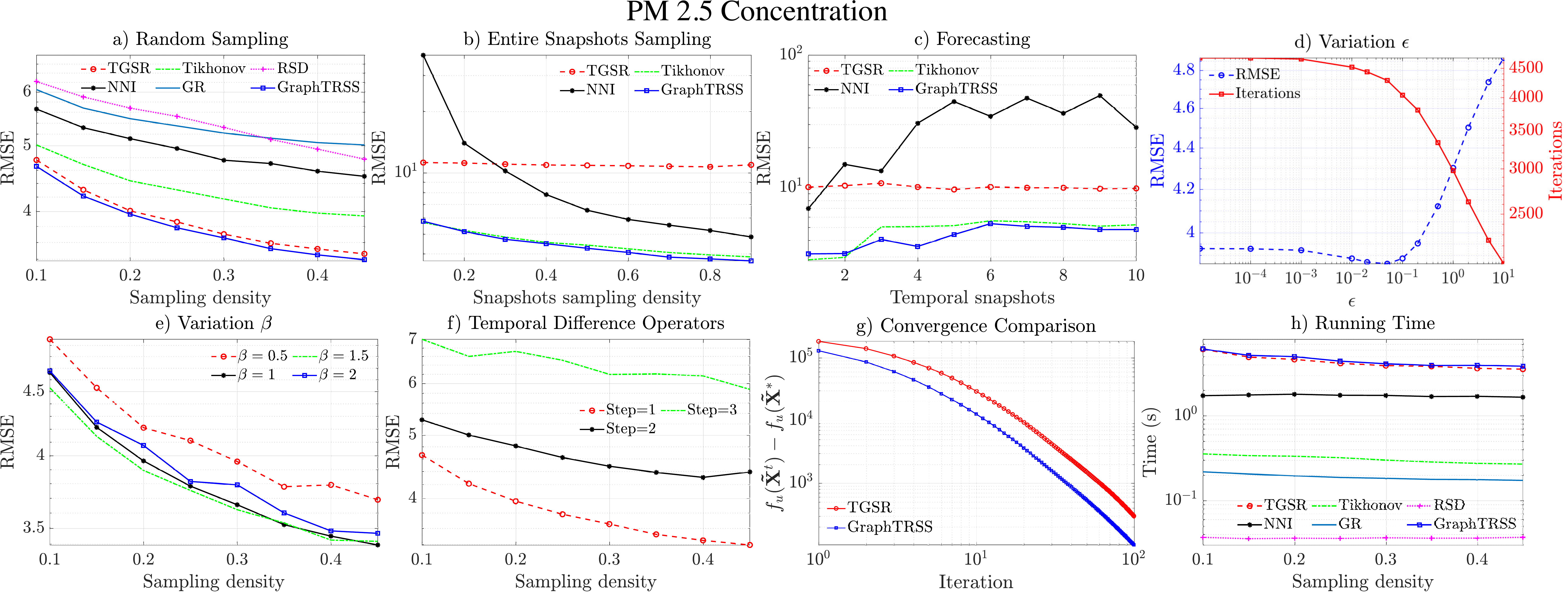}
    \caption{Comparison of GraphTRSS with several methods in the literature in the PM 2.5 dataset for several experiments in terms: a) random sampling, b) entire snapshots sampling, c) forecasting, d) variation of parameter $\epsilon$, e) variation of parameter $\beta$, f) several temporal difference operators, g) convergence comparison, and h) running time.}
    \label{fig:results_pm_25}
\end{figure*}

\begin{figure*}
    \centering
    \includegraphics[width=\textwidth]{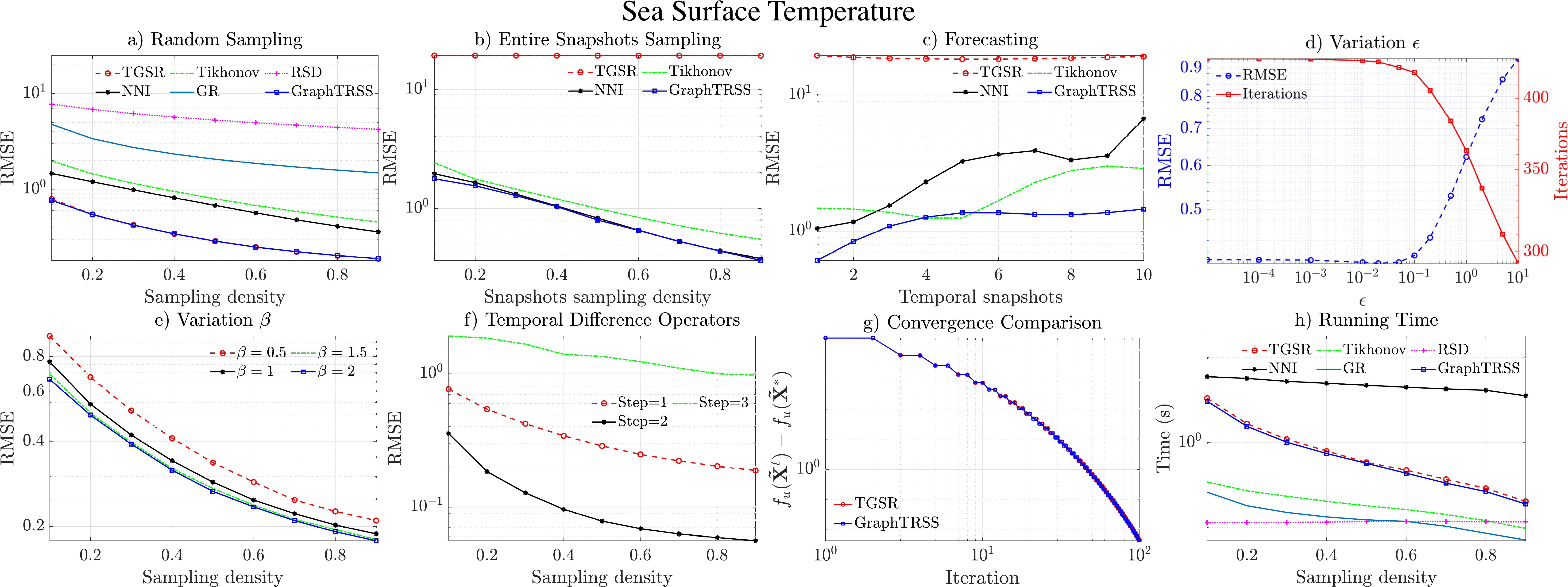}
    \caption{Comparison of GraphTRSS with several methods in the literature in the sea surface temperature dataset for several experiments in terms of: a) random sampling, b) entire snapshots sampling, c) forecasting, d) variation of parameter $\epsilon$, e) variation of parameter $\beta$, f) several temporal difference operators, g) convergence comparison, and h) running time.}
    \label{fig:results_sea}
\end{figure*}

\subsection{COVID-19 Datasets}

Fig. \ref{fig:results_covid_global} and \ref{fig:results_covid_USA} show the results of the experiments in the COVID-19 datasets.
GraphTRSS is better than the other methods for the reconstruction experiments when using different sampling strategies.
We can notice that the results in the COVID-19 USA dataset have lower RMSE than the global dataset, \ie we can get better estimations if we have more fine-grained data as in the case of the COVID-19 USA dataset.
From the experiment with entire temporal snapshots, we can notice that TGSR method \cite{qiu2017time} performs poorly compared to the other methods for what we observe in random sampling.
This behavior is expected because entire snapshots sampling does not satisfy the second condition of Theorem \ref{trm:uniqueness_solution} for TGSR by Qiu \etal \cite{qiu2017time}.

All tested methods, including ours, are designed to reconstruct time-varying graph signals, and therefore these methods do not have prior assumptions about the behavior in the time domain of the underlying processes.
Correspondingly, in forecasting, we get weaker results than the performance shown in Fig. \ref{fig:results_covid_global} and \ref{fig:results_covid_USA} for random and entire snapshots sampling.
However, our algorithm readily shows better results than the other methods in both COVID-19 datasets, even forecasting multiple time steps ahead.
Thus, the Sobolev function introduced in Definition \ref{dfn:sobolev_norm_time_varying} could potentially improve other graph-based forecasting methods of the literature \cite{isufi2019forecasting}, using proper time-domain prior assumptions or other machine learning strategies, which we leave for future work.

From the studies to analyze the parameters of GraphTRSS, one can notice that values of $\epsilon>0$ improve the condition number; this is reflected in the number of iterations to satisfy the stopping condition as shown in Fig. \ref{fig:results_covid_global}(d) and \ref{fig:results_covid_USA}(d) for COVID-19.
However, one should be careful since big values of $\epsilon$ can heavily modify the structure of the graph leading to a degradation of the performance, as shown in Fig. \ref{fig:results_covid_global}(d) and \ref{fig:results_covid_USA}(d) for the RMSE.
For the variations of $\beta$ and the temporal difference operators, we notice that values different from $\beta=1$ and one time-step for the temporal difference operator might bring benefits for COVID-19 datasets.

Fig. \ref{fig:results_covid_global}(g) shows that GraphTRSS converges faster than TGSR method for COVID-19 global, where $\mathbf{\tilde{X}}^*$ is the solution of the optimization problem, and $\mathbf{\tilde{X}}^t$ is the solution at iteration $t$.
Fig. \ref{fig:results_covid_global}(h) and \ref{fig:results_covid_USA}(h) show the running time for each method.
These experiments were computed with the best parameters of each method from the random sampling experiment.
Arguably, GraphTRSS shows the best compromise between accuracy and running time among several methods for COVID-19 datasets, as shown in Fig. \ref{fig:results_covid_global} and \ref{fig:results_covid_USA}.

\subsection{Environmental Datasets}

Fig. \ref{fig:results_pm_25} and \ref{fig:results_sea} show the results for the experiments in the environmental datasets.
The analysis of the comparison between our algorithm and the other methods is mostly similar to the analysis for Fig. \ref{fig:results_covid_global} and \ref{fig:results_covid_USA}, \ie GraphTRSS shows the best compromise between accuracy and running time.
However, we should notice two interesting results in Fig. \ref{fig:results_pm_25} and \ref{fig:results_sea}.
Firstly, we notice that the parameters $\beta=1.5$ or $\beta=2$ show better results than $\beta=1$ in the environmental datasets.
Secondly, the two-time steps temporal operator $\mathbf{D}_{h2}$ shows a big performance improvement compared to the one-step temporal operator for the sea surface temperature dataset.
Perhaps, we could propose other temporal operators and more elaborated optimization functions based on several $\beta$ parameters, which we leave for future work.

\begin{table}
\centering
\caption{The average number of iterations to satisfy the stopping condition for GraphTRSS and TGSR \cite{qiu2017time}.}
\label{tbl:number_iterations}
\makebox[\linewidth]{
\scalebox{0.77}{
\begin{tabular}{l|cccc}
\hline
Method    & COVID-19 Global & COVID-19 USA & PM 2.5 Conc. & Sea Surface Temp. \\
\hline
TGSR \cite{qiu2017time}       & 1735.9          & 48659       & 4690.1               & 418.8                   \\
GraphTRSS & 510.0           & 427         & 4266.6               & 416.0 \\
\hline
\end{tabular}
}
}
\end{table}

\begin{figure}
    \centering
    \includegraphics[width=0.48\textwidth]{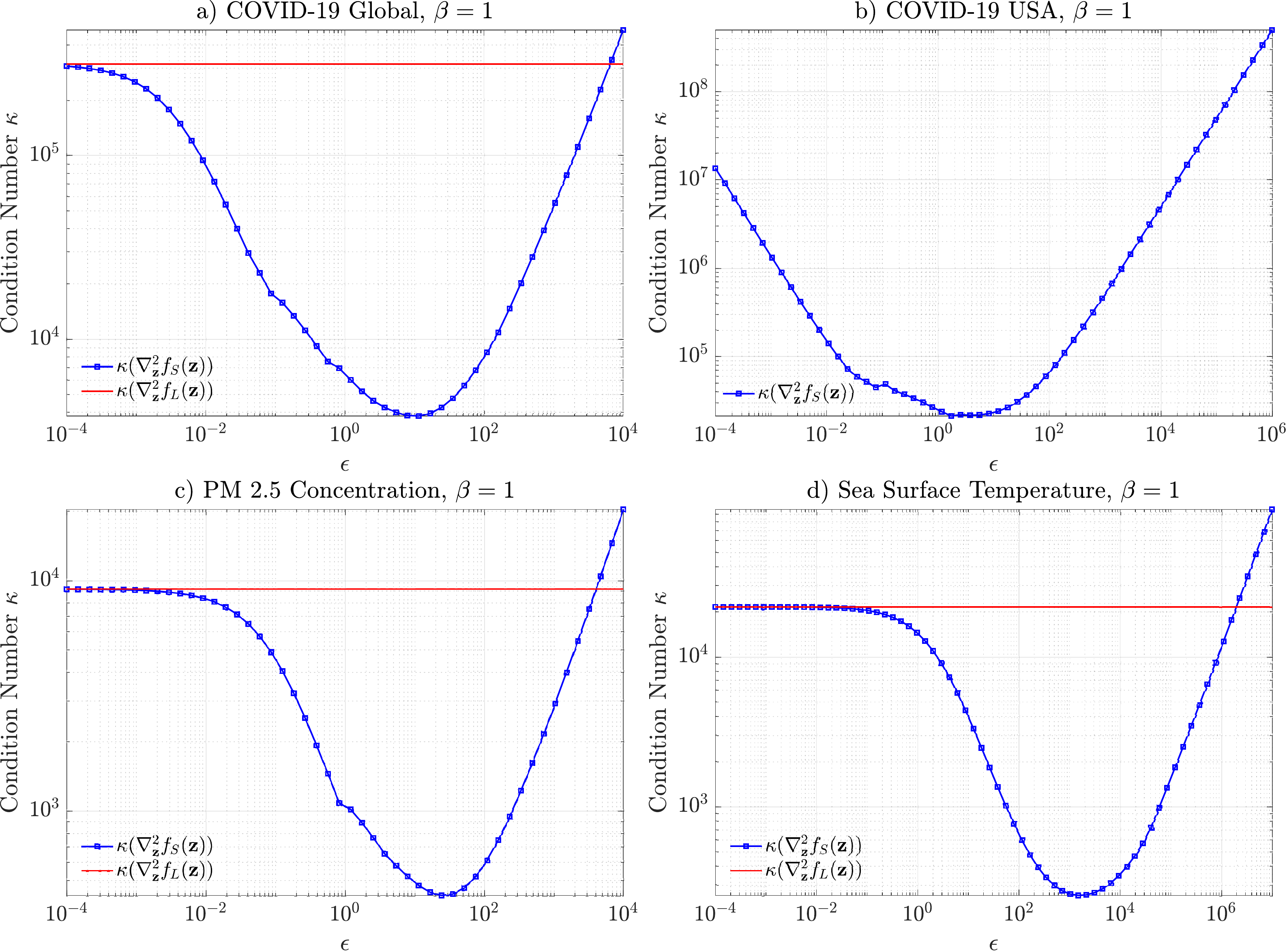}
    \caption{Condition number of the Hessian associated with the optimization problems for GraphTRSS ($\kappa(\nabla_{\mathbf{z}}^2 f_{S}(\mathbf{z}))$) and TGSR \cite{qiu2017time} ($\kappa(\nabla_{\mathbf{z}}^2 f_{L}(\mathbf{z}))$) for all datasets.}
    \label{fig:condition_number}
\end{figure}

\subsection{Additional Analysis}

Table \ref{tbl:number_iterations} shows the number of iterations that TGSR method and GraphTRSS require to satisfy the stopping condition in \eqref{eqn:stopping_condition}.
Table \ref{tbl:number_iterations} shows that GraphTRSS satisfies the stopping condition around $3.4$ times faster than TGSR in global COVID-19; and about $114$ times faster in USA COVID-19.
We limit the maximum number of iterations of TGSR to $60000$ for the USA COVID-19 dataset because of time limitations.
The experiments with the lowest sampling densities would take more time to satisfy the stopping condition if we increase the maximum number of iterations for TGSR method in COVID-19 USA.
This behavior is another undesirable effect of bad condition numbers.

Fig. \ref{fig:condition_number} shows the condition number of the Hessian functions for the COVID and environmental datasets with random sampling with density $0.5$.
The condition number of the Hessian of \eqref{eqn:qiu_reconstruction_noisy} is not displayed in Fig. \ref{fig:condition_number} for COVID-19 USA because, in that case, $\kappa(\nabla_{\mathbf{z}}^2 f_{L}(\mathbf{z})) \to \infty$.
Furthermore, as predicted by Theorem \ref{trm:convergence_rate}: 1) there is a range of values of $\epsilon$ where $\kappa(\nabla_{\mathbf{z}}^2 f_{S}(\mathbf{z})) < \kappa(\nabla_{\mathbf{z}}^2 f_{L}(\mathbf{z}))$ and then we get benefits in convergence rate, and 2) $\kappa(\nabla_{\mathbf{z}}^2 f_{S}(\mathbf{z}))$ grows quickly for large values of $\epsilon$.
Notice that the improvement in the condition number depends on the specific structure of each graph.
Therefore, using the normalized Laplacian $\mathbf{\mathcal{L}}=\mathbf{D}^{-\frac{1}{2}}\mathbf{LD}^{-\frac{1}{2}}$ is reasonable since $\lambda_{max}(\mathbf{\mathcal{L}}) \leq 2$.
The interested readers are referred to Section VI in the supplementary material for a small experiment with normalized Laplacian matrices.

The parameters $\epsilon$ and ${\upsilon}$ are optimized for each dataset to get good performance.
However, we can notice in Fig. \ref{fig:condition_number} that an $\epsilon$ value close to $10^{-1}$ is a good starting point since it improves the condition number, and it does not modify heavily the Laplacian matrix.
Similar analysis about the parameter ${\upsilon}$ can be found in \cite{qiu2017time}.
In practice, the parameters of GraphTRSS can be optimized in a small part of the dataset.

\section{Conclusions}
\label{sec:conclusions}

In this work, a new algorithm called GraphTRSS is introduced for the reconstruction of time-varying graph signals.
The Sobolev norm is used to define a smoothness function for time-varying graph signals, and therefore this function is used to introduce a new optimization problem for reconstruction from samples.
We showed the good convergence properties of GraphTRSS by relying on the condition number of the Hessian associated with our problem, as shown in Theorem \ref{trm:convergence_rate}.
Our algorithm improves convergence rate without relying on expensive matrix inversions or eigenvalue decompositions.
GraphTRSS shows promising results for the estimation of new COVID-19 cases in both global and USA datasets, as well as in the reconstruction of environmental variables (PM 2.5 concentration and sea surface temperature).
GraphTRSS could be useful in several scenarios, for example, 1) one could try to estimate the number of new COVID-19 cases where we have missing or unreliable information, 2) we can use our algorithm as a pre-processing step to get more reliable data for further forecasting of new cases with other well-established models of infectious diseases \cite{hethcote2000mathematics}, and 3) we can also use GraphTRSS as a pre-processing step for further forecasting of new COVID-19 cases with other graph-based techniques such as \cite{isufi2019forecasting}.
GraphTRSS is also evaluated directly on forecasting, showing promising results.

This work shows several insights for possible future directions.
For example, our algorithm and other GSP techniques could potentially increase the performance of various models in infectious disease (not only for COVID-19) for forecasting and imputation of data, among others.
In the same way, this work could be the foundation for applying other GSP-based methods in virology.
These methods can be, for example, graph neural networks \cite{kipf2017semi,bronstein2017geometric} and learning graphs from data \cite{dong2019learning}.

\appendices
\section{Proof Theorem \ref{trm:convergence_rate}}
\label{app:proof_convergence}

\begin{proof}
    The problem associated with the Laplacian matrix in (\ref{eqn:qiu_reconstruction_noisy}) can be rewritten as follows \cite{qiu2017time}:
    \begin{equation}
        f_L(\mathbf{z})=\min_{\mathbf{z}} \frac{1}{2} \Vert \mathbf{Q}[\mathbf{z}-\vectorized{(\mathbf{Y})}] \Vert_2^2+\frac{{\upsilon}}{2}\mathbf{z}^{\mathsf{T}}[(\mathbf{D}_h\mathbf{D}_h^{\mathsf{T}})\otimes \mathbf{L}] \mathbf{z},
        \label{eqn:qiu_reconstruction_noisy_rewritten}
    \end{equation}
    where $\mathbf{z}=\vectorized{(\mathbf{\tilde{X}})}$.
    Moreover, the gradient of $f_L(\mathbf{z})$ in (\ref{eqn:qiu_reconstruction_noisy_rewritten}) is such that:
    \begin{equation}
        \nabla_{\mathbf{z}} f_L(\mathbf{z}) = \mathbf{Q}[\mathbf{z}-\vectorized{(\mathbf{Y})}]+{\upsilon} [(\mathbf{D}_h\mathbf{D}_h^{\mathsf{T}})\otimes \mathbf{L}]\mathbf{z},
    \end{equation}
    and the Hessian matrix of $f_L(\mathbf{z})$ is given as follows:
    \begin{equation}
        \nabla_{\mathbf{z}}^2 f_L(\mathbf{z}) = \mathbf{Q} + [{\upsilon} (\mathbf{D}_h\mathbf{D}_h^{\mathsf{T}})\otimes \mathbf{L}].
        \label{eqn:hessian_quadratic_form}
    \end{equation}
    Similarly, the Hessian matrix associated with the Sobolev formulation is given by:
    \begin{equation}
        \nabla_{\mathbf{z}}^2 f_S(\mathbf{z}) = \mathbf{Q} + [{\upsilon} (\mathbf{D}_h\mathbf{D}_h^{\mathsf{T}})\otimes (\mathbf{L}+\epsilon\mathbf{I})^{\beta}].
        \label{eqn:hessian_sobolev}
    \end{equation}
    $\mathbf{D}_h\mathbf{D}_h^{\mathsf{T}}$ is a positive semi-definite matrix, so it has a matrix of eigenvalues $\mathbf{\Lambda}_D=\diag(\lambda_{(D)1},\lambda_{(D)2},\dots,\lambda_{(D)N})$, with $0 \leq \lambda_{(D)1} \leq \lambda_{(D)2} \leq \dots \leq \lambda_{(D)N}$, and the corresponding matrix of eigenvectors $\mathbf{U}_D$.
    As a consequence, from \eqref{eqn:hessian_quadratic_form}, we have:
    \begin{gather}
        \nonumber
        \nabla_{\mathbf{z}}^2 f_L(\mathbf{z}) = \mathbf{Q} + [{\upsilon} (\mathbf{U}_D\mathbf{\Lambda}_D\mathbf{U}_D^{\mathsf{T}}) \otimes (\mathbf{U}\mathbf{\Lambda}\mathbf{U}^{\mathsf{T}})],\\
        \label{eqn:Hessian_laplacian_final}
        \nabla_{\mathbf{z}}^2 f_L(\mathbf{z}) = {\upsilon} \left[ \frac{1}{{\upsilon}} \mathbf{Q} + (\mathbf{U}_D \otimes \mathbf{U}) (\mathbf{\Lambda}_D \otimes \mathbf{\Lambda}) (\mathbf{U}_D^{\mathsf{T}} \otimes \mathbf{U}^{\mathsf{T}})\right],
    \end{gather}
    where we used the property of Kronecker products $(\mathbf{A} \otimes \mathbf{B})(\mathbf{C} \otimes \mathbf{D}) = \mathbf{AC} \otimes \mathbf{BD}$ \cite{horn2012matrix}.
    $\mathbf{Q}$ is positive semi-definite because it is a diagonal matrix with diagonal elements either $0$ or $1$.
    As a result, from \eqref{eqn:Hessian_laplacian_final}, we know that $\lambda_{min}(\frac{1}{{\upsilon}} \mathbf{Q}) = 0$, $\lambda_{max}(\frac{1}{{\upsilon}} \mathbf{Q}) = \frac{1}{{\upsilon}}$ if $\mathbf{J} \neq \mathbf{0}$, $\lambda_{min}((\mathbf{D}_h\mathbf{D}_h^{\mathsf{T}})\otimes \mathbf{L}) = 0$, and $\lambda_{max}((\mathbf{D}_h\mathbf{D}_h^{\mathsf{T}})\otimes \mathbf{L}) = \lambda_N \lambda_{(D)N}$ if $\lambda_N,\lambda_{(D)N} \geq 1$.
    For the Sobolev problem, we have that:
    \begin{equation}
        (\mathbf{L}+\epsilon\mathbf{I})^{\beta} = (\mathbf{U\Lambda U}^{\mathsf{T}}+\epsilon\mathbf{I})^{\beta} = \mathbf{U}(\mathbf{\Lambda}+\epsilon\mathbf{I})^{\beta}\mathbf{U}^{\mathsf{T}},
    \end{equation}
    and then:
    \begin{equation}
        \label{eqn:Hessian_Sobolev_final}
        \resizebox{\hsize}{!}{
        $\nabla_{\mathbf{z}}^2 f_{S}(\mathbf{z}) = {\upsilon} \left[ \frac{1}{{\upsilon}} \mathbf{Q} + (\mathbf{U}_D \otimes \mathbf{U}) (\mathbf{\Lambda}_D \otimes (\mathbf{\Lambda}+\epsilon \mathbf{I})^{\beta}) (\mathbf{U}_D^{\mathsf{T}} \otimes \mathbf{U}^{\mathsf{T}})\right],$
        }
    \end{equation}
    where $\lambda_{min}((\mathbf{D}_h\mathbf{D}_h^{\mathsf{T}})\otimes (\mathbf{L}+\epsilon \mathbf{I})^{\beta}) = 0$ and $\lambda_{max}((\mathbf{D}_h\mathbf{D}_h^{\mathsf{T}})\otimes (\mathbf{L}+\epsilon \mathbf{I})^{\beta}) = (\lambda_N + \epsilon)^{\beta} \lambda_{(D)N}$ if $(\lambda_N + \epsilon)^{\beta},\lambda_{(D)N} \geq 1$.
    We have a summation of two Hermitian matrices in \eqref{eqn:Hessian_laplacian_final} and \eqref{eqn:Hessian_Sobolev_final} (we can trivially prove that \eqref{eqn:Hessian_laplacian_final} and \eqref{eqn:Hessian_Sobolev_final} are Hermitian matrices since $(\mathbf{A} \otimes \mathbf{B})^{\mathsf{T}}=\mathbf{A}^{\mathsf{T}} \otimes \mathbf{B}^{\mathsf{T}}$).
    Therefore, we cannot get an exact value for the eigenvalues of these Hessian functions.
    The problem of describing the possible eigenvalues of the sum of two Hermitian matrices in terms of their spectra has attracted the interest of mathematicians for decades.
    The most complete description was conjectured by Horn \etal \cite{horn1962eigenvalues}.
    The interested readers are referred to \cite{knutson2001honeycombs} for an interesting discussion in the topic.
    However, we can get some bounds using the developments of Weyl \cite{horn2012matrix}:
    \begin{theorem}[Weyl's Theorem (4.3.1 in \cite{horn2012matrix})]
        \label{trm:weyl_theorem}
        Let $\mathbf{A}$ and $\mathbf{B}$ be Hermitian matrices with set of eigenvalues $\{a_1,a_2,\dots,a_N\}$ and $\{b_{1},b_{2},\dots,b_{N}\}$, respectively, where the eigenvalues are in ascending order.
        Then, $c_i \leq a_{i+j} + b_{N-j}$, with $j=0,1,\dots,N-i$, and $a_{i-j+1}+b_j \leq c_i$, with $j=1,\dots,i$, where $c_i$ is the $i$th eigenvalue of $\mathbf{A}+\mathbf{B}$.\\
        Proof: see \cite{horn2012matrix}.
    \end{theorem}
    Using Theorem \ref{trm:weyl_theorem} and \eqref{eqn:Hessian_laplacian_final}, we can have the following inequalities for $\nabla_{\mathbf{z}}^2 f_{L}(\mathbf{z})$:
    \begin{gather}
        \label{eqn:lambda_max_laplacian}
        \lambda_N \lambda_{(D)N} \leq \lambda_{max}(\nabla_{\mathbf{z}}^2 f_{L}(\mathbf{z})) \leq \lambda_N \lambda_{(D)N} + \frac{1}{{\upsilon}},\\
        \label{eqn:lambda_min_lambda_Laplacian}
        0 \leq \lambda_{min}(\nabla_{\mathbf{z}}^2 f_{L}(\mathbf{z})) \leq \frac{1}{{\upsilon}}, \\
        \label{eqn:lambda_min_Laplacian}
        0 \leq \lambda_{min}(\nabla_{\mathbf{z}}^2 f_{L}(\mathbf{z})) \leq \lambda_N \lambda_{(D)N}.
    \end{gather}
    Similarly, for $\nabla_{\mathbf{z}}^2 f_{S}(\mathbf{z})$ we have:
    \begin{gather}
        \nonumber
        (\lambda_N + \epsilon)^{\beta} \lambda_{(D)N} \leq \lambda_{max}(\nabla_{\mathbf{z}}^2 f_{S}(\mathbf{z}))\\
        \label{eqn:lambda_max_epsilon}
        \leq (\lambda_N + \epsilon)^{\beta} \lambda_{(D)N} + \frac{1}{{\upsilon}},\\
        \label{eqn:lambda_min_lambda_Sobolev}
        0 \leq \lambda_{min}(\nabla_{\mathbf{z}}^2 f_{S}(\mathbf{z})) \leq \frac{1}{{\upsilon}}, \\
        \label{eqn:lambda_min_Sobolev}
        0 \leq \lambda_{min}(\nabla_{\mathbf{z}}^2 f_{S}(\mathbf{z})) \leq (\lambda_N + \epsilon)^{\beta} \lambda_{(D)N},
    \end{gather}
    Notice that we ignored the ${\upsilon}$ value that multiplies \eqref{eqn:Hessian_laplacian_final} and \eqref{eqn:Hessian_Sobolev_final} in these inequalities since $\kappa({\upsilon} \mathbf{A}) = \kappa(\mathbf{A})$.
    We can have some asymptotic analysis based on the inequalities \eqref{eqn:lambda_max_laplacian}-\eqref{eqn:lambda_min_Sobolev}:
    \begin{enumerate}[leftmargin=*]
        \item For $\epsilon > 0$ and $\beta > 1$ we have that the upper bound in \eqref{eqn:lambda_min_Sobolev} is looser than the upper bound in \eqref{eqn:lambda_min_Laplacian} given that $\lambda_N+\epsilon \geq 1$, which favors better condition numbers for the Sobolev problem.
        For example, if we assume that the eigenvalues are equal to the upper bounds in \eqref{eqn:lambda_max_laplacian}, \eqref{eqn:lambda_min_Laplacian}, \eqref{eqn:lambda_max_epsilon}, and \eqref{eqn:lambda_min_Sobolev} we have that:
        \begin{gather}
            \nonumber
            \kappa(\nabla_{\mathbf{z}}^2 f_{S}(\mathbf{z})) = \left( 1+\frac{1}{{\upsilon}(\lambda_N + \epsilon)^{\beta} \lambda_{(D)N}} \right)^2 \\
            \leq \kappa(\nabla_{\mathbf{z}}^2 f_{L}(\mathbf{z})) = \left( 1+\frac{1}{{\upsilon}\lambda_N \lambda_{(D)N}}\right)^2.
        \end{gather}
        \item When ${\upsilon} \to \infty$, $\lambda_{min}(\nabla_{\mathbf{z}}^2 f_{L}(\mathbf{z})) \to 0$ and $\lambda_{min}(\nabla_{\mathbf{z}}^2 f_{S}(\mathbf{z})) \to 0$ according to \eqref{eqn:lambda_min_lambda_Laplacian} and \eqref{eqn:lambda_min_lambda_Sobolev}, and then $\kappa(\nabla_{\mathbf{z}}^2 f_{L}(\mathbf{z})) \to \infty$ and $\kappa(\nabla_{\mathbf{z}}^2 f_{S}(\mathbf{z})) \to \infty$.
        \item When $\epsilon \to \infty$ and $\beta>0$, $\lambda_{max}(\nabla_{\mathbf{z}}^2 f_{S}(\mathbf{z})) \to \infty$ according to \eqref{eqn:lambda_max_epsilon} and then $\kappa(\nabla_{\mathbf{z}}^2 f_{S}(\mathbf{z})) \to \infty$.
        When $\epsilon \to \infty$ and $\lambda_{min}(\nabla_{\mathbf{z}}^2 f_{L}(\mathbf{z})) > 0$ we have that $\kappa(\nabla_{\mathbf{z}}^2 f_{S}(\mathbf{z})) > \kappa(\nabla_{\mathbf{z}}^2 f_{L}(\mathbf{z}))$.
        \item when $\beta \to \infty$ and $\lambda_N+\epsilon > 1$, $\lambda_{max}(\nabla_{\mathbf{z}}^2 f_{S}(\mathbf{z})) \to \infty$ according to \eqref{eqn:lambda_max_epsilon} and then $\kappa(\nabla_{\mathbf{z}}^2 f_{S}(\mathbf{z})) \to \infty$.
    \end{enumerate}
\end{proof}

%\section*{Acknowledgment}

%The authors thank the support provided by Consejo Nacional de Ciencia y Tecnolog\'ia (CONACYT) Mexico under the grant number 770750 to carry out this research work.

\ifCLASSOPTIONcaptionsoff
  \newpage
\fi



%\bibliographystyle{IEEEtran}

\bibliographystyle{IEEEtran}
\bibliography{bibfile}


\begin{IEEEbiography}[{\includegraphics[width=1in,height=1.25in,clip,keepaspectratio]{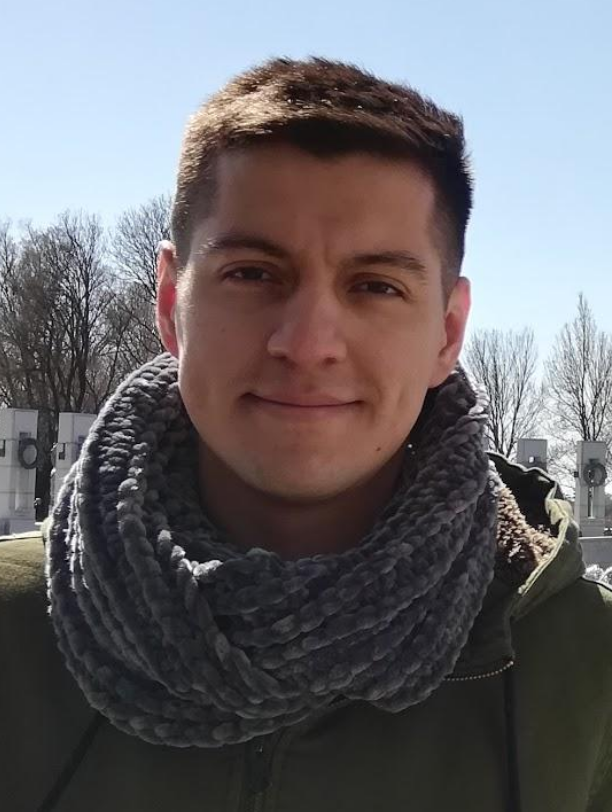}}]{Jhony H. Giraldo}
received the B.Sc. and M.Sc. degrees in Electronics Engineering from the University of Antioquia, Medellin, Colombia. He is currently working toward a Ph.D. degree in Computer Sciences with the University of La Rochelle, France. His research interests include the fundamentals and applications of computer vision, machine learning, graph signal processing, and graph neural networks. He has worked on image and video processing, supervised and semi-supervised learning, and on sampling and reconstruction of graph signals.%, and the modeling of infectious diseases using concepts of graph theory and signal processing.
\end{IEEEbiography}

%\newpage

\begin{IEEEbiography}[{\includegraphics[width=1in,height=1.25in,clip,keepaspectratio]{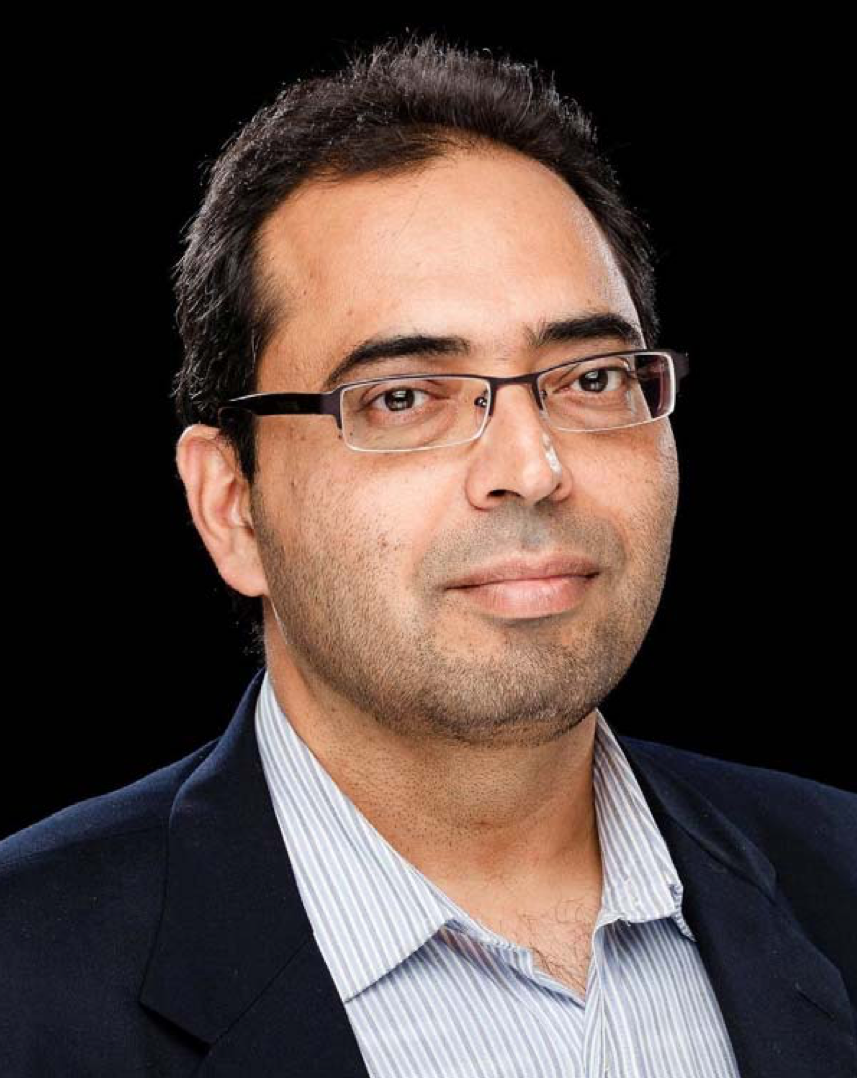}}]{Arif Mahmood}
is currently a Professor with Information Technology University (ITU). He received his M.Sc. and Ph.D. in Computer Science from the Lahore University of Management Sciences (LUMS). He has served as a Research Assistant Professor at the School of Computer Science and Software Engineering, and later at the School of Mathematics and Statistics in the University of Western Australia from 2012-2015. Previously, he was an Assistant Professor at the Punjab University College of Information Technology from 2008-2012. He also worked as a Postdoctoral Researcher at the College of Engineering in Qatar University from 2015-2018. His major research interests include data clustering, classification, action and object recognition using image sets. He has also worked on computation elimination algorithms for fast template matching, video compression, object removal and image mosaicking. At the School of Mathematics and Statistics, he worked on community detection in complex networks.
\end{IEEEbiography}

\begin{IEEEbiography}[{\includegraphics[width=1in,height=1.25in,clip,keepaspectratio]{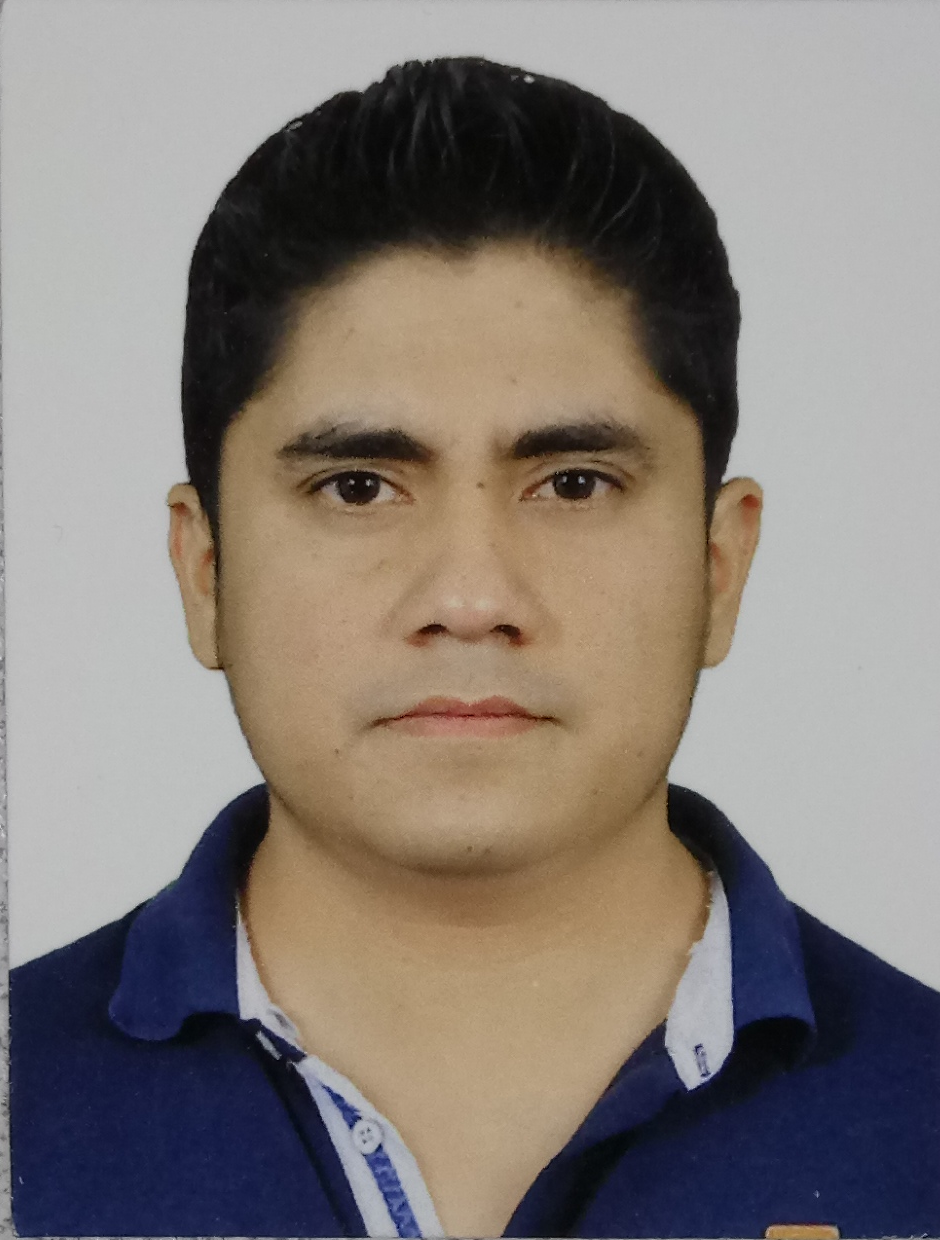}}]{Belmar Garcia-Garcia}
received the B.Sc. degree in Communications and Electronics Engineering, the M.Sc. degree in Electronics Engineering, and the Ph.D. degree in Communications and Electronics Engineering from Instituto Politécnico Nacional de México (IPN). He is currently a Postdoctoral Researcher at the University of La Rochelle, France. He is focused on the development of algorithms for prediction and diagnosis of COVID-19. Other topics of interest are tracking, digital electronic, real time processing, and artificial intelligence.
\end{IEEEbiography}

\begin{IEEEbiography}[{\includegraphics[width=1in,height=1.25in,clip,keepaspectratio]{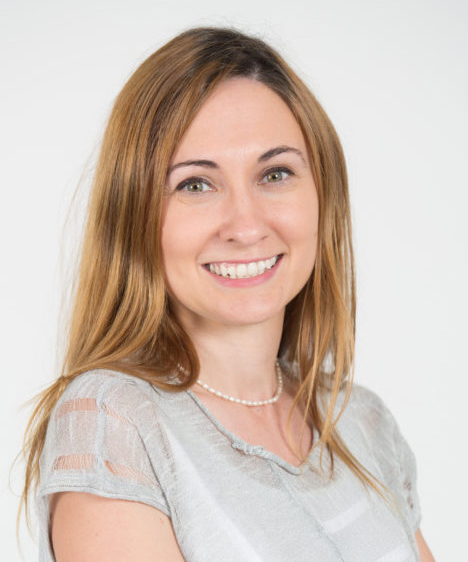}}]{Dorina Thanou}
is a senior researcher and lecturer at EPFL, leading the development of the Intelligent Systems for Medicine and Health research pillar, under the Center for Intelligent Systems. Prior to that, she was a Senior Data Scientist at the Swiss Data Science Centre. She got her M.Sc. and Ph.D. in Communication Systems and Electrical Engineering respectively, both from EPFL, Switzerland, and her Diploma in Electrical and Computer Engineering from the University of Patras, Greece. She is the recipient of the Best Student Paper Award at ICASSP 2015, and the co-author of the Best Paper Award at PCS 2016. She is an Ellis Scholar. Her research interests lie in the broader area of signal processing and machine learning, with a special focus on graph-based representation learning algorithms. She is particularly interested in applying her expertise to intelligent systems for healthcare.
%received the B.Sc. degree in electrical and computer engineering from the University of Patras, Greece, in 2008, and the M.Sc. degree in communication systems and the Ph.D. degree in electrical engineering from the Swiss Federal Institute of Technology (EPFL), Lausanne, in 2010 and 2016, respectively. She was a Postdoctoral Researcher with the Signal Processing Laboratory, EPFL and a Senior Data Scientist with the Swiss Data Science Centre (SDSC), EPFL/ETH Zurich. She is currently the lead of the Intelligent Systems for Medicine and Health Program at CIS-EPFL since November 2020. Her research interests include graph-based signal processing for data representation and analysis as well as machine learning, with a particular focus on the design of interpretable models for real-world applications.
\end{IEEEbiography}

\begin{IEEEbiography}[{\includegraphics[width=1in,height=1.25in,clip,keepaspectratio]{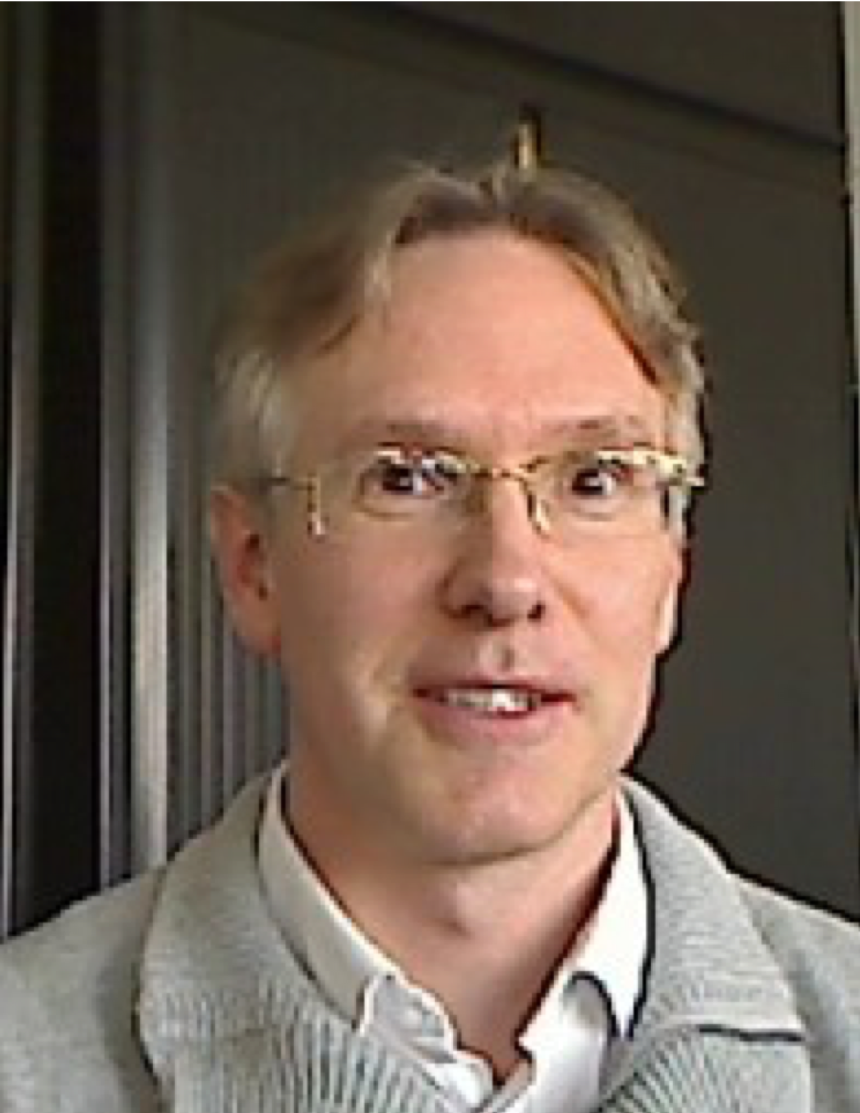}}]{Thierry Bouwmans}
is an Associate Professor at the University of La Rochelle, France. His research interests consist mainly in the detection of moving objects in challenging environments. He has recently authored more than 30 papers in the field of background modeling and foreground detection. These papers investigated particularly the use of fuzzy concepts, discriminative subspace learning models and robust PCA. They also develop surveys on mathematical tools used in the field and particularly on Robust PCA via Principal Component Analysis. He has supervised Ph.D. students in this field. He is the creator and administrator of the Background Subtraction Web Site. He served as the lead guest editors in two editorial works: 1) the special issue in MVA on Background Modeling for Foreground Detection in Real-World Dynamic Scenes and 2) the Handbook on "Background Modeling and Foreground Detection for Video Surveillance" in CRC press. He has served as a reviewer for numerous international conferences and journals.
\end{IEEEbiography}


%\vfill


\end{document}